\documentclass[sigconf,10pt]{acmart}

\AtBeginDocument{%
  }

\setcopyright{acmlicensed}
\copyrightyear{2018}
\acmYear{2018}
\acmDOI{XXXXXXX.XXXXXXX}

\acmConference[Conference acronym 'XX]{Make sure to enter the correct conference title from your rights confirmation emai}{June 03--05, 2018}{Woodstock, NY}

\acmISBN{978-1-4503-XXXX-X/18/06}

\usepackage{balance}
\usepackage{amsmath}               
\usepackage{caption}
\usepackage{subcaption}
\usepackage{float}
\usepackage{wrapfig}
\usepackage{listings}
\usepackage[frozencache=true,cachedir=minted-cache]{minted}
\usepackage{lipsum}
\usepackage{parcolumns}
\usepackage{paralist}
\usepackage{algorithm}
\usepackage[noend]{algpseudocode}

\usepackage{adjustbox}
\usepackage{multirow}
\usepackage{pifont}
\usepackage{eqparbox}
\usepackage{enumitem}
\setlist[itemize]{noitemsep, topsep=0pt, itemindent=0pt,
leftmargin=\parindent}

\definecolor{applegreen}{rgb}{0.55, 0.71, 0.0}

\algnewcommand\algorithmicforeach{\textbf{for each}}
\algdef{S}[FOR]{ForEach}[1]{\algorithmicforeach\ #1\ \algorithmicdo}

\usepackage{xparse}
\makeatletter
\NewDocumentCommand{\LeftComment}{s m}{%
  \Statex \IfBooleanF{#1}{\hspace*{\ALG@thistlm}}\(\triangleright\) #2}
\makeatother

\usepackage{color}
\definecolor{codegreen}{rgb}{0,0.6,0}
\definecolor{codegray}{rgb}{0.5,0.5,0.5}
\definecolor{codepurple}{rgb}{0.58,0,0.82}
\definecolor{backcolour}{rgb}{0.95,0.95,0.92}

\begin{document}

\copyrightyear{2024} 
\acmYear{2024} 
\setcopyright{rightsretained} 
\acmConference[SoCC '24]{ACM Symposium on Cloud Computing}{November 20--22, 2024}{Redmond, WA, USA}
\acmBooktitle{ACM Symposium on Cloud Computing (SoCC '24), November 20--22, 2024, Redmond, WA, USA}
\acmDOI{10.1145/3698038.3698540}
\acmISBN{979-8-4007-1286-9/24/11}

\title{Rethinking State Management in Actor Systems for Cloud-Native Applications}

\author{Yijian Liu}
\affiliation{
\institution{University of Copenhagen}
\country{Denmark}
}
\email{liu@di.ku.dk}

\author{Rodrigo Laigner}
\affiliation{
\institution{University of Copenhagen}
\country{Denmark}
}
\email{rnl@di.ku.dk}

\author{Yongluan Zhou}
\affiliation{
\institution{University of Copenhagen}
\country{Denmark}
}
\email{zhou@di.ku.dk}

\begin{abstract}
%The actor model has emerged as an effective solution for building stateful middle tiers. By promoting the partitioning of application functionalities and associated states into actors, developers benefit from their inherent concurrency model to achieve scalability and isolation. 
%This gap challenges developers seeking such requirements and inhibits further adoption of actor systems in data-intensive applications.

The actor model has gained increasing popularity. However, it lacks support for complex state management tasks, such as enforcing foreign key constraints and ensuring data replication consistency across actors. These are crucial properties in partitioned application designs, such as microservices. %which inhibit the further adoption of the actor model. 
To fill this gap, we start by analyzing the key impediments in state-of-the-art actor systems. We find it difficult for developers to express complex data relationships across actors and reason about the impact of state updates on performance due to opaque state management abstractions. 
To solve this conundrum, we develop \texttt{SmSa}, a novel data management layer for actor systems, allowing developers to declare data dependencies that cut across actors, including foreign keys, data replications, and other dependencies. \texttt{SmSa} can transparently enforce the declared dependencies, reducing the burden on developers. Furthermore, \texttt{SmSa} employs novel logging and concurrency control algorithms to support transactional maintenance of data dependencies. 

We demonstrate \texttt{SmSa} can support core data management tasks where dependencies across components appear frequently without jeopardizing application logic expressiveness and performance. Our experiments show \texttt{SmSa} significantly reduces the logging overhead and leads to increased concurrency level, improving by up to 2X the performance of state-of-the-art deterministic scheduling approaches. As a result, \texttt{SmSa} will make it easier to design and implement highly partitioned and distributed applications.
\end{abstract}





\maketitle

\section{Introduction}

In the modern application programming landscape, developers design their applications in tiered architectures with a stateless middle tier encoding the application logic and a database tier storing the application state~\cite{Bernstein2014}. For each client request, the middle tier executes the business logic by retrieving the necessary data from the database tier. This architecture simplifies the development of applications by pushing complex data management tasks to the database tier. However, this data shipping paradigm has limitations. The long end-to-end latency and excessive data transfer from the database to the middle tier during peak periods may not meet the applications' requirements~\cite{AODB, Shah2017}. In-memory data caching in the middle tier can reduce the latency and data transfer. However, it can suffer from low cache hit ratios, e.g., when different client requests need to access diverse data or data are updated frequently, and cache data inconsistency that can lead to application safety problems~\cite{ports12}.

To address these problems, an alternative architecture employs a stateful middle tier, where data are stored in the computing nodes, and the function shipping paradigm is adopted. Client requests are shipped to the computing nodes, which store the corresponding data for processing. This does not involve data transfer from the data storage, minimizing the bandwidth overhead and latency. Data from the middle tier are asynchronously shipped to the database layer for analytics and disaster recovery. Furthermore, to achieve system scalability, software agility, and fault isolation, we are witnessing the emergence of microservice architectures~\cite{vldb2021}. In such architectures, the application is decomposed into independent and fine-grained components that interact via synchronous or asynchronous communications, each encapsulating its own state. 

Meanwhile, the actor model~\cite{Agha1986} has emerged as a promising concurrent programming model for middle-tier development~\cite{WhyAkka}. The encapsulated states of actors allow developers to build loosely coupled applications with a design that remounts a cache with data locality, characterizing the function shipping paradigm. Client requests are processed by one or more actors interacting via asynchronous messages, triggering the updates of their encapsulated actor states. %As a result of the function, a message can be generated for asynchronous processing of another actor, allowing developers to compose actors’ behaviour to achieve certain functionalities. 
Such design principle makes it very attractive to model microservices as actors~\cite{Marketplace}. 
In addition, the advent of virtual actor~\cite{Bernstein2014}, originally developed in the context of Microsoft Orleans, further alleviates developers' burden by providing actor state management functionalities, including state persistence, transactional state manipulation, and fault tolerance via state persistence, which lays a solid foundation for addressing the data management challenges in microservice systems~\cite{vldb2021}. The state of a virtual actor is modeled as an opaque transactional object, no matter how many entities are encapsulated in an actor. Such an opaque state model puts almost no limitation on how developers implement the operations on actor states other than using the required APIs to interact with an underlying data store for persistence. Thus, an update in a single entity is treated as an update in the object as a whole.

%, it may potentially jeopardize transaction performance due to the coarse-grained state operations. 
\begin{figure}[tb]
    \centering
    \includegraphics[width=\linewidth]{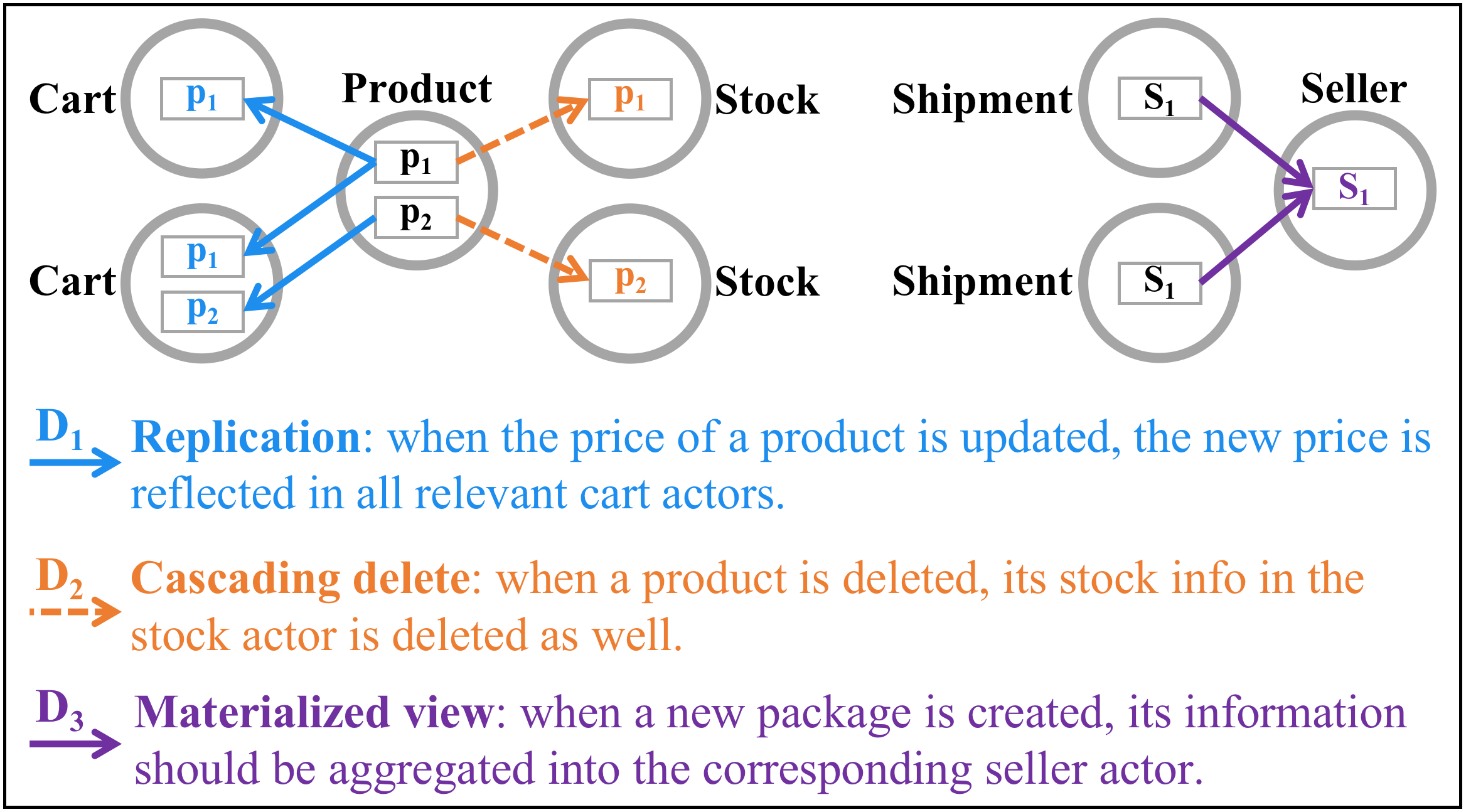}
    \caption{Cross-actor relations in \texttt{Online MarketPlace}}
    \label{fig:example}
\end{figure}
Despite the recent advancement of actor state management, modern data-intensive applications exhibit complex relationships among entities cutting across distributed components~\cite{synapse}, which typically are not natively supported by actor runtimes. These cross-component relationships include foreign-key constraints, replication of data items, and functional dependencies~\cite{vldb2021}. Taking an e-commerce application case, \texttt{Online Marketplace}~\cite{laigner2024benchmarking}, as an example (Fig.\ref{fig:example}), product data managed by the product component can be replicated to the cart component, which manages products added to customer carts. The replication favors performance since it avoids successive round trips between the cart and product components for ensuring the correctness of product information (e.g., product price) on checkout time~\cite{synapse,quote_intro}. However, the opaque state model of virtual actors exposes little semantic information, such as how and which part of the state is modified, thereby limiting the ability of developers to express important safety properties of data-intensive applications. Due to the lack of support from the actor frameworks, developers must map relationships across actors explicitly (e.g., which carts contain a certain product) and ensure their correctness in application code. Besides being complex and error-prone, this practice is oblivious to transaction management, leading to isolation anomalies. 

Taken together, enhancing the state management features of actor frameworks for ensuring application safety across actors is a missing key to fully realizing the envisioned benefits of the actor model in developing scalable stateful middle tiers. To bridge the gap, we develop a data management library for virtual actors, which provides an API for developers to register cross-actor dependency constraints dynamically and supports high-throughput transactional enforcement of these constraints. It accounts for emerging cross-component state management requirements that arise in partitioned and distributed applications, such as microservices~\cite{vldb2021, olep, synapse}. Our contributions are summarized as follows:

\begin{itemize}
    \item To strike a balance between simplicity, state translucency, and expressiveness, we propose an extended key-value model for actor states with an associated set of state access and dependency APIs. This design enables the shift of cross-actor dependency management from the application codes to the actor framework and hence unloads the burden of developers from complex and error-prone state management codes.
    \item We develop \texttt{SmSa}, a data management layer in a virtual actor framework. \texttt{SmSa} is implemented by integrating our data model and APIs with the state-of-the-art transaction library for virtual actors -- \texttt{Snapper} to support transactional data dependency enforcement. To maximize the transaction throughput, we extend \texttt{Snapper}'s concurrency control to take advantage of the new data model of actor states, providing fine-grained concurrency control outperforming vanilla \texttt{Snapper}.
    \item We map the \texttt{Online Marketplace}~\cite{laigner2024benchmarking} benchmark to the actor abstraction and implement it on \texttt{SmSa}. \texttt{Online Marketplace} models key safety properties related to relationships across components that arise in real-world partitioned and distributed applications.  Our implementation validates the expressiveness and ease of use of our data model, as well as addresses the challenges of cache coherence, referential integrity, and strong isolation transactional guarantee.
    \item We conduct extensive experiments on two benchmarks, \texttt{Online Marketplace} and \texttt{SmallBank}. Our experiment demonstrates the efficiency of fine-grained concurrency control over state-of-the-art deterministic methods. As a result, \texttt{SmSa} will facilitate the design and implementation of actor-based stateful middle-tiers.
\end{itemize}
\section{Actor State Management for Data Integrity}

%In this section, we present \texttt{SmSa}, a novel actor state management layer that allows developers to declare data dependencies that cut across actors, including foreign keys, data replications, and other dependencies. We start by equipping actors with an explicit data model to aid developers in managing actor states
%based on application entities rather than opaque state objects  ($\S$~\ref{subsec:data_model}). 
%Next, we describe how we leverage an actor data model to fulfil key data dependencies that arise in modern partitioned and distributed applications ($\S$~\ref{sec:dependency}). We conclude by devising novel logging and concurrency control algorithms to support transactional maintenance of data dependencies transparently to developers ($\S$~\ref{subsec:tx_actor}). Taken together, these advances make the actor model more attractive to a wider range of developers working on stateful middle-tier applications.

% In this section, we introduce \texttt{SmSa}, an actor state management layer built for actor-based applications. \texttt{SmSa} aims to unload a variety of state management tasks, including maintaining data replication consistency and enforcing data integrity constraints, from application codes to system-level support through a dedicated data model.
\subsection{Orleans}
Orleans~\cite{Orleans} is a framework that facilitates the development of distributed applications through virtual actors~\cite{Bernstein2014}. Orleans automatically allocates virtual actors in available nodes and deactivates actors when they are no longer needed. In addition, Orleans transparently migrates virtual actors from faulty computational resources to others in case of failures, ensuring that the application remains functional.

To aid developers in designing data-intensive applications, Orleans recently introduced state management and transaction management APIs~\cite{OrleansTxn} (\texttt{OrleansTxn}). Through transactional objects of user-defined types as the state of actors~\cite{tx_state}, Orleans can preclude developers from explicitly implementing transactional guarantees. \texttt{OrleansTxn} adopts a locking-based concurrency control method and applies Early-Lock Release to achieve a higher concurrency level. However, as reported by recent research \cite{Snapper, Marketplace}, \texttt{OrleansTxn} is vulnerable to contentions, often leading to low performance. 

\begin{figure*}[h]
    \centering
    \includegraphics[width=\linewidth]{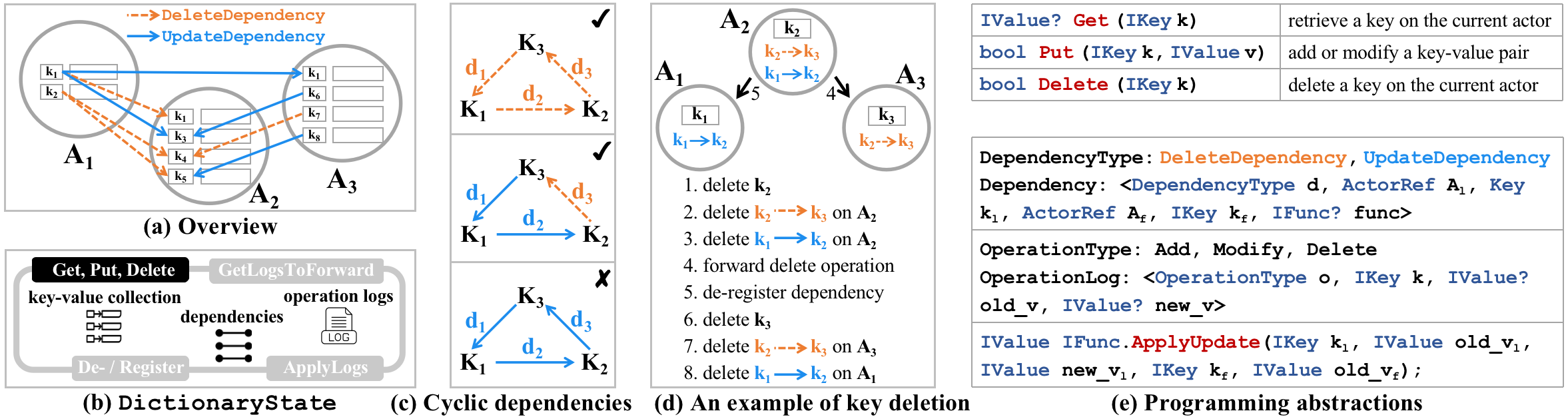}
    \caption{State management of stateful actors (\texttt{SmSa})}
    \label{fig:model}
\end{figure*}

\subsection{Conceptual Overview of \texttt{SmSa}}
\label{subsec:data_model}
In \texttt{SmSa}, each actor's state is modeled as a key-value collection. \texttt{SmSa} relies on developers to determine how keys are partitioned across actors. Each key is unique among those in an actor, while the same key may exist in different actors. Keys located in different actors are related through a dependency constraint ($\S$~\ref{sec:dependency}). As shown in Fig.\ref{fig:model}a, actors and dependencies form a directed graph. The arrow $k_i \rightarrow k_j$ represents that $k_j$ depends on $k_i$, and a change made on $k_i$ may cause a change to $k_j$. \texttt{SmSa} supports basic operations to get, put, and delete keys; meanwhile, it allows dynamic registration and de-registration of dependencies between two specific pairs of key-value items in two different actors. Thus, the dependency graph may change with time, well reflecting the dynamic nature of actor topologies~\cite{orleans_streams}. 
%This differs from relational database systems, where references between entities and table schemes need to be declared before declaring references between different attributes. 
Besides, we opt for a key-value abstraction because developers are familiar with this model to manage data in stateful middle-tiers~\cite{ports12,mertz}.

The dependencies in \texttt{SmSa} are used to model different application constraints, such as data integrity constraints, foreign key constraints, data replication, and other functional dependencies. For example, Fig.\ref{fig:example} illustrates the replicated key (\textbf{D$_1$}), the foreign key with cascading delete (\textbf{D$_2$}), and the materialized view (\textbf{D$_3$}), respectively. A prominent feature of \texttt{SmSa} is the system-level support to enforce these constraints instead of relying on developers to encode them case by case in the application logic. More specifically, \texttt{SmSa} keeps track of the operations performed on the keys during the execution of the actor methods. When the execution is done, \texttt{SmSa} scans the list of operation logs and figures out the operations that need to be forwarded to the other keys on the other actors for preserving the relevant dependency constraints. For example, in Fig.\ref{fig:example}, the dependency \textbf{D$_1$} represents a scenario when the price of a product is changed, the update operation is forwarded to all cart actors that contain this product, thus guaranteeing that all the cart actors consistently see the latest price. With \texttt{SmSa}, the forwarded update operations are calculated and carried out transparently at the system level. $\S$~\ref{sec:dependency} discusses the algorithm applied to derive such information and how operations are forwarded and applied to another actor.

Fig.\ref{fig:model}b illustrates that each actor maintains not only a key-value collection but also a list of dependencies and a list of operation logs. \texttt{SmSa} wraps them together into a \texttt{DictionaryState} object. This object exposes two sets of APIs; one is external for the user request to access application data, and the other is internal for the system to enforce the dependency constraints. By enforcing certain operations being invoked via these APIs, \texttt{SmSa} captures all the changes made to the key-value collection of an actor and further automates the dependency constraint enforcement.

\subsection{Dependency Management}
\label{sec:dependency}

\subsubsection{Definition}
\texttt{SmSa} defines a dependency record with six data fields (Fig.\ref{fig:model}e): the type of the dependency, the leader actor, the leader key, the follower actor, the follower key, and a customized function. \texttt{SmSa} generalizes two types of dependencies -- delete dependency and update dependency. The \textbf{delete dependency} describes the relation between two keys ($k_l$, $k_f$) on two different actors ($A_l$, $A_f$) that the deletion of $k_l$ on $A_l$ will cause the deletion of $k_f$ on $A_f$. The \textbf{update dependency} refers to the relation in which the update that happened to $k_l$ on $A_l$ will trigger a specified update function being applied to the key $k_f$ on $A_f$. In an update dependency, the deletion of $k_l$ will not cause the deletion of $k_f$. A key where the delete or update operation originates is named \textbf{leader key}, the affected key is called \textbf{follower key}, and the actors where the keys are located are referred to as \textbf{leader actor} and \textbf{follower actor} respectively. The leader key and the follower key do not have to be the same, depending on how users define the dependency. To be able to retrieve the follower keys of a leader key, \texttt{SmSa} stores the dependency information on the leader actor. \texttt{SmSa} does not maintain a shared global dependency graph; instead, it distributes the information across actors. By doing so, \texttt{SmSa} avoids a centralized component becoming a bottleneck when it needs to serve frequent queries, meanwhile guaranteeing that each actor has sufficient information stored in the local private state to enforce the dependencies.

\begin{figure*}
    \centering
    \includegraphics[width=\linewidth]{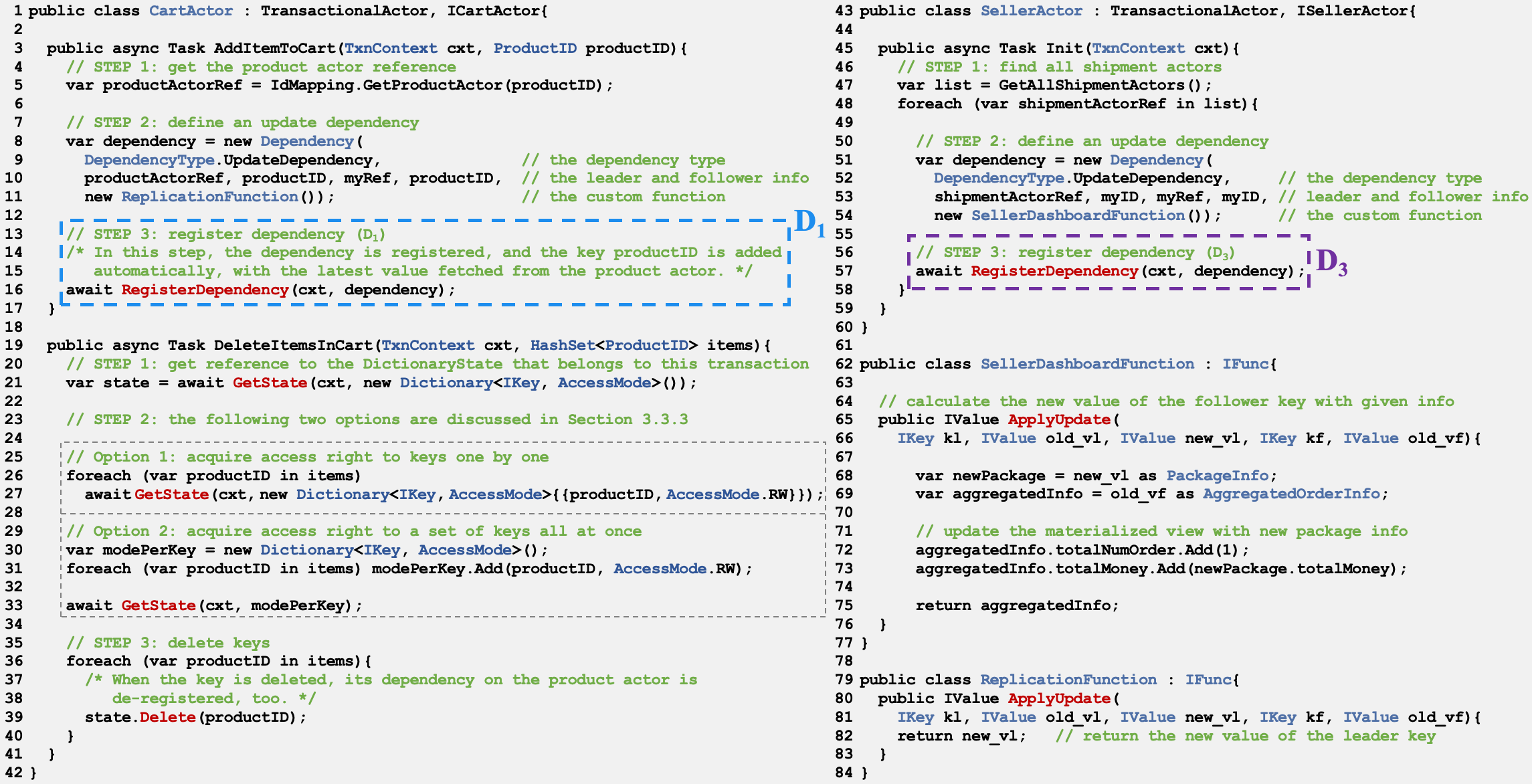}
    \caption{Example codes of developing \texttt{Online Marketplace} on \texttt{SmSa-X}}
    \label{fig:code}
\end{figure*}

In addition, an update function needs to be specified for every update dependency. \texttt{SmSa} defines an interface $IFunc$ and an abstract method $ApplyUpdate$ (Fig.\ref{fig:model}e) that each update function should implement. This method requires input: the leader key, the values of the leader key before and after applying the user-invoked update operation, the follower key, and the existing value of the follower key. Then, the method returns the calculated new value for the follower key. As is shown in Fig.\ref{fig:code} (lines 79-84), data replication can be facilitated by using an update function that directly returns the value of the leader key. Thus, the follower key remains the same as the leader. This function can be used for defining dependency \textbf{D$_1$} where the new price of a product is replicated from the product actor to all relevant cart actors (lines 3-17). Similarly, customized functions can contain more complex calculations and be applied to build materialized views. As the example in Fig.\ref{fig:code} (lines 62-77), a function is defined for a seller actor to maintain a view of all created orders cumulatively.

\subsubsection{Deletion Rules of Keys}
\label{sec:deletion}
\texttt{SmSa} allows user code to delete both leader and follower keys. When a leader key is deleted, the deletion operation is forwarded to the corresponding followers. When a follower key is deleted, it indicates its dependency on the leader key expires; thus, this dependency record should be deleted to avoid further operations being sent to the follower actor (e.g., lines 35-40 in Fig.\ref{fig:code}). In summary, the deletion of a key indicates the deletion of follower keys, as well as the de-registration of all dependencies pointing to and from this key. To retrieve the dependency in the backward direction, i.e., from the follower key to the leader key, \texttt{SmSa} keeps a copy of the dependency record on the follower actor as well.

\subsubsection{Leader and Follower Keys}
In \texttt{SmSa}, a leader key may have multiple follower keys. It resembles the scenario in an RDBMS where multiple foreign keys refer to one primary key. \texttt{SmSa} allows a leader key to impose different effects on different follower keys. For example, in Fig.\ref{fig:example}, a product ID $p_1$ on the product actor may relate to multiple cart actors through update dependencies (\textbf{D$_1$}). Meanwhile, $p_1$ is also related to the same key on the stock actor through a delete dependency (\textbf{D$_3$}).

%For example, in Fig.\ref{fig:model}a, a leader key $k_1$ on the actor $A_1$ may relate to a key $k_1$ on $A_2$ through a delete dependency while relating to keys $k_3$ on $A_2$ and $k_1$ on $A_3$ through update dependencies. The key $k_1$ on $A_3$ can be used to model a replica where the replicated information remains consistent with the original copy. The key $k_3$ on $A_2$ models a functional dependency on $k_1$ on $A_1$.

\texttt{SmSa} allows a follower key to have multiple leader keys, while a foreign key can only reference a single primary key in RDBMS. \texttt{SmSa} identifies three scenarios. First, a follower key may have update dependencies on multiple leader keys. For example, as is shown in Fig.\ref{fig:example}, a follower key $s_1$ on the seller actor may need to aggregate the information of newly created orders from the leader keys on multiple shipment actors. By doing so, the seller actor can maintain an up-to-date materialized view. Second, a follower key may have delete dependencies on multiple leader keys, which models the case that the existence of an entity is dependent upon the existence of other entities. Third, a follower key may be affected by both update and delete operations on different leader keys. Note that \texttt{SmSa} does not control the priority or the order of executing operations related to different dependencies. When a delete operation and an update operation are both forwarded to the same follower key, the final result is the same -- the follower key is deleted. However, when two different update operations are forwarded to the same follower key, the result may vary depending on which update operation arrives at the follower actor first or which update function is applied by the follower actor first. Users can resolve this issue by defining commutative update functions. Another solution is to rely on transaction management ($\S$~\ref{subsec:tx_actor}) to restrict the order of different operations performed on each actor.

% when a key is both leader and follower
In \texttt{SmSa}, when a key acts both as a leader and a follower, the deletion of such a key becomes more complicated. For example, in Fig.\ref{fig:model}d, when the key $k_2$ is deleted, the actor $A_2$ first enquiries forwards to send the delete operation to the follower actor $A_3$, then backwards to de-register the dependency on $A_1$. 

\subsubsection{Cyclic Dependencies}
\label{sec:cycle}
If dependency registration is not restricted, \texttt{SmSa} may create a cyclic chain of dependencies across multiple keys, which will further result in an infinite loop of operation forwarding process. As is shown in Fig.\ref{fig:model}c, the cycle may be formed in three cases. The first case involves only delete dependencies, the second case is a mix of both types of dependencies, and the third one only has update dependencies. 

The first two can be processed to completion within limited steps. In the first case, the deletion of the key $k_1$ will cause the deletion of $k_2$, then $k_3$. When the deletion operation is forwarded back to the actor where $k_1$ is located, the actor will find out that $k_1$ does not exist in the key-value collection because it has already been deleted. Thus, the deletion will not be performed again, and no more delete operations will be forwarded along the dependency chain. As for the second case, the update of $k_3$ will be forwarded to $k_1$ and $k_2$, and it does not form a cycle. When $k_1$ ($k_3$) is deleted, the two connected dependencies $d_1$ and $d_2$ ($d_3$ and $d_1$) are removed, and no further operations are needed. When $k_2$ is deleted, $k_3$, $d_2$, $d_3$ and $d_1$ will be deleted. 

In the third case, deleting any of the keys will cause the deletion of two dependencies. However, updating any of the keys will cause the update operation to circulate endlessly in the cycle. The third type of cyclic dependencies can be prohibited by adopting the following three strategies: \textbf{S$_1$}, \texttt{SmSa} does additional checks whenever an update dependency is registered, which may be time-consuming when there exists a very long chain of dependency, or when each key points to many other keys. \textbf{S$_2$}, \texttt{SmSa} can restrict that each actor should not forward operations for a request more than once. For example, if an update is already forwarded from $k_1$ to $k_2$, then when $k_1$ receives the update from $k_3$, no more updates will be generated. However, this strategy requires actors to store information about different requests, introduces extra semantics, and makes it more complicated to reason about the application logic. \textbf{S$_3$}, \texttt{SmSa} can disallow a key to becoming both leader and follower, thus eliminating any type of cycle.

\subsubsection{Dependency Registration and De-registration}
\label{sec:register-dependency}
In \texttt{SmSa}, a dependency registration request should be initiated by the user on the follower actor $A_f$. $A_f$ first checks if the specified $k_f$ is allowed to be attached with a dependency. For example, \texttt{SmSa} may check if it will cause a cyclic chain of update dependency when the strategy S$_1$ is adopted, or check if $k_f$ is already identified as a leader when S$_3$ is adopted. If the check is passed, the leader actor $A_l$ is called to continue the registration. $A_l$ will check if the key $k_l$ can be declared as a leader, i.e. if $k_l$ already exists and there is no dependency that has $k_l$ as a follower key. If yes, the dependency is registered on $A_l$, and the latest value $v_l$ of the leader key $k_l$ is retrieved and returned to $A_f$. On $A_f$, if the follower key $k_f$ does not exist yet, $v_l$ is used as the initial value of $k_f$, and a new key-value pair $<k_f, v_l>$ is inserted. If $k_f$ already exists, the custom function specified in the newly registered dependency is applied by using $old\_v_l = v_l$ and $new\_v_l = v_l$. The above explains how \texttt{SmSa} deals with a dependency registration request. Next, we explain how developers should send such a request to \texttt{SmSa}. Fig.\ref{fig:code} shows the example codes of registering dependencies \textbf{D$_1$} (lines 3-17) and \textbf{D$_3$} (lines 45-59). In both cases, a dependency is defined by specifying the six required data fields. Afterward, the API $RegisterDependency$ is called to forward the necessary information to internal \texttt{SmSa}. \texttt{SmSa} exposes this API for developers to explicitly declare dependencies.

The de-registration of a dependency can be carried out in two ways. First, it can be done in a similar process as dependency registration, where a de-registration request is sent explicitly to the follower actor and then forwarded to the leader actor. Second, the user can also simply delete the follower key (Fig.\ref{fig:code} lines 35-40). According to the deletion rule of \texttt{SmSa} introduced in $\S$~\ref{sec:deletion}, when a follower key is deleted, the corresponding dependency records are removed as well.

\begin{figure*}[h]
    \centering
    \includegraphics[width=\linewidth]{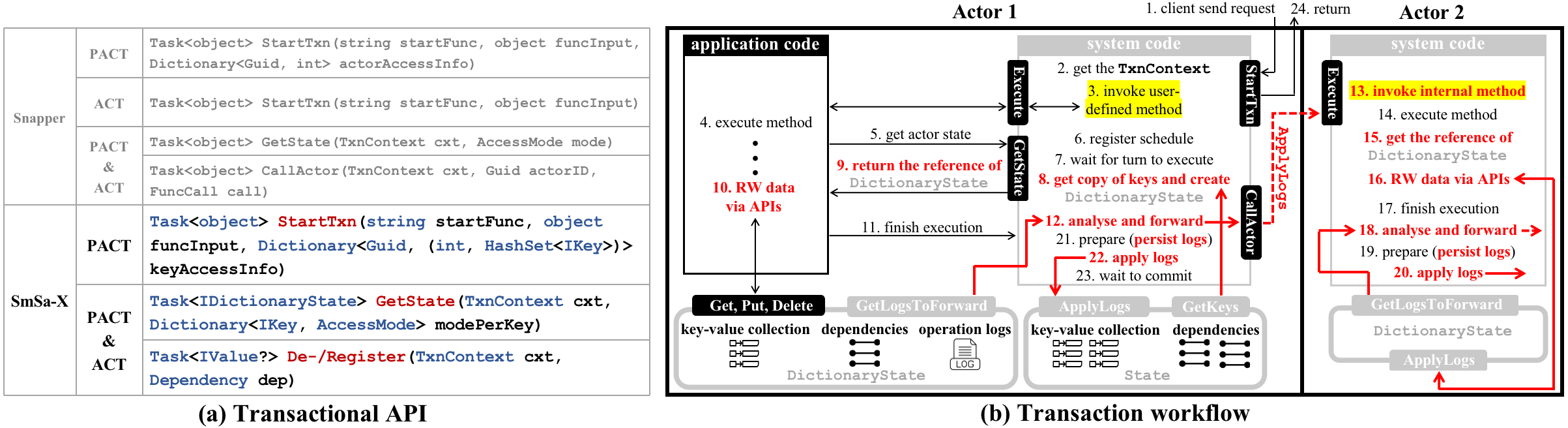}
    \caption{Integration of \texttt{SmSa} and \texttt{Snapper} -- \texttt{SmSa-X}}
    \label{fig:workflow}
\end{figure*}
\section{Transactional Actor State Management}
\label{subsec:tx_actor}

Having laid out a rich state management abstraction for the actor model, we turn our attention to how to leverage such advancements to optimize key aspects of data systems, logging and concurrency control.
% \texttt{SmSa} provides an abstraction to manage each actor's state as a key-value collection which allows developers to manipulate the actor state in a fine-grained manner. \texttt{SmSa} also generalizes various application constraints as dependency relations across actors and provides methods for developers to identify and declare these constraints. By introducing the key-value data model, the dependency model, as well as the data structure to record changes, \texttt{SmSa} automates the enforcement of dependency constraints and further alleviates the burden of developers.

\begin{wrapfigure}{l}{0.18\textwidth}
    \vspace{-3ex}
    \centering
    \includegraphics[scale=0.24]{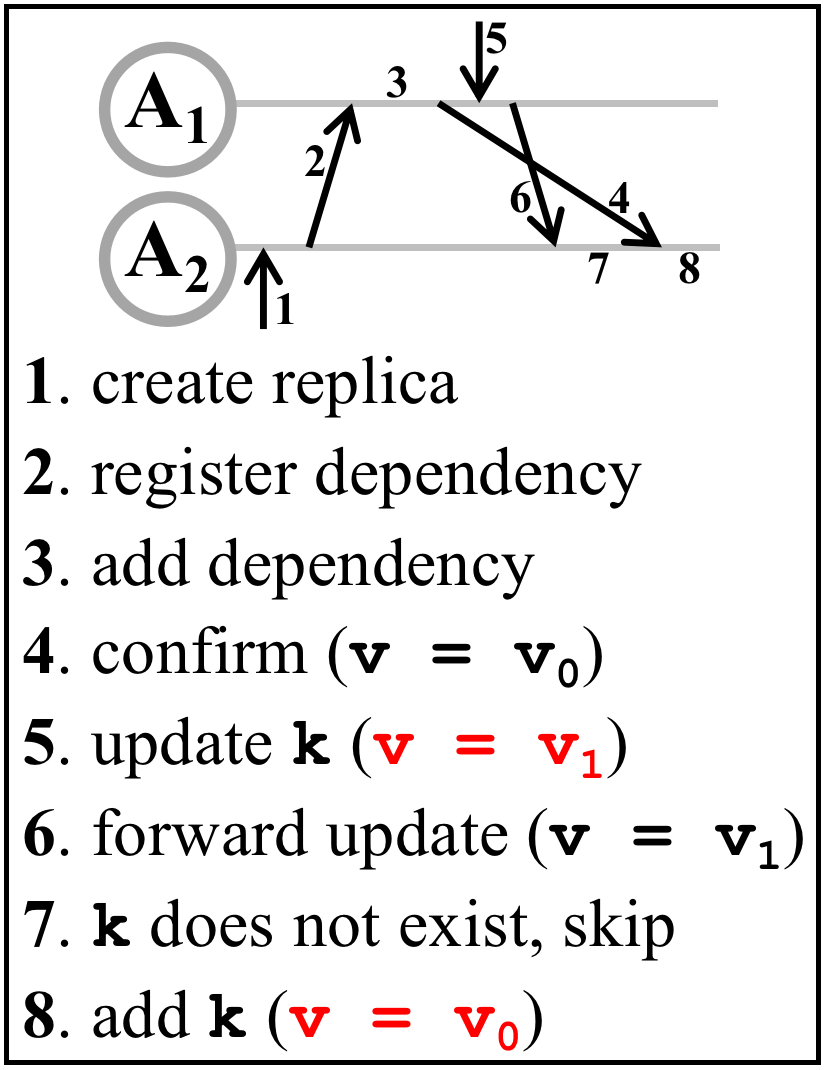}
    \vspace{-5ex}
    \caption{}
    \label{fig:consistency}
    \vspace{-4ex}
\end{wrapfigure}
\subsection{The Dangers of Unordered Operations}
Although equipping actors with a data model enables the declaration of complex relationships across actors, developers must still account for the dangers of arbitrary function execution order and their possible impact on actor states~\cite{actor_bugs}. For instance, a request in \texttt{SmSa} may perform operations on multiple actors, especially in the existence of cross-actor dependencies. The concurrent execution of such requests may drive the system to an inconsistent state. For example, in Fig.\ref{fig:consistency}, suppose an actor $A_2$ intends to create a replica of key $k$, whose master copy is stored on $A_1$. The example illustrates a possible interleaving of two requests; one is a dependency registration request initiated by $A_2$, and the other performs an update operation on $k$ on $A_1$. Ideally, the updated value of key $k$ should be reflected to the replica on $A_2$. However, if the forwarded update operation arrives earlier than the confirmation message, $A_2$ will not perform the update and results in a replica that is inconsistent with the master copy on $A_1$.

\texttt{SmSa} can benefit from a transaction management solution to provide stronger application correctness guarantees. Existing works have proposed a myriad of methods to enforce the order of concurrent tasks \cite{Snapper, OrleansTxn} and to converge to a consistent state in the presence of disorder \cite{anna,Preguica2018,CRDT, Akka-CRDT}. \texttt{Snapper} \cite{Snapper}, one of these solutions, supports ACID transactional properties for multi-actor operations through performant deterministic concurrency control, making it a good fit to integrate \texttt{SmSa} with.

% In this section, we take advantage of \texttt{Snapper}'s coherent transaction programming abstraction, an actor-based architecture, as well as its performant transaction processing techniques. More specifically, we integrate \texttt{SmSa} with \texttt{Snapper} to achieve transactional guarantees on dependencies that cut across actors.

\subsection{An Actor Transaction Library -- \texttt{Snapper}}
\subsubsection{Transaction Management}
\texttt{Snapper}~\cite{Snapper} is an actor transaction library designed to enhance the performance of multi-actor transactions meanwhile preserving the ACID transactional properties. \texttt{Snapper} leverages deterministic scheduling for PACT, a type of transaction that declares the actor access information when it is submitted to the system. Based on this pre-declared information, \texttt{Snapper} employs a group of coordinator actors to generate deterministic transaction execution schedules for every batch of PACTs. Each such schedule consists of a sequence of PACTs that are relevant to a particular actor. Afterwards, \texttt{Snapper} delivers such batch schedules to corresponding actors via asynchronous messages. Each actor is supposed to execute transactional invocations of PACTs one by one in the pre-determined order. Under the PACT execution, transactions are scheduled, executed, committed, and logged all at the batch level.

In addition to deterministic execution, \texttt{Snapper} also supports conventional non-deterministic strategies, such as Strict Two-Phase Locking (S2PL) and Two-Phase  Commit (2PC), for transactions whose actor access information cannot be pre-declared. This type of transaction is called ACT. Another standout feature of Snapper is its ability to execute concurrent hybrid workloads, wherein some transactions are executed deterministically while others follow non-deterministic methods. This hybrid execution model maximizes the advantages of deterministic execution, which can achieve high transaction throughput while maintaining flexibility for non-deterministic workloads.

\subsubsection{Programming Model}
%\texttt{Snapper} is an actor transaction library built on top of Orleans that provides a high-level API for efficient scheduling and execution of deterministic and non-deterministic multi-actor transactions, namely PACT and ACT, respectively. 
\texttt{Snapper} provides a base actor class, \texttt{TransactionalActor}, that provides key system-level functionalities, such as transaction processing and logging. As listed in Fig.\ref{fig:workflow}a, the \texttt{TransactionalActor} exposes APIs for transaction requests submission (\texttt{StartTxn}), for transactional actor state access (\texttt{GetState}), and for invoking transactional methods on other actors (\texttt{CallActors}).

In \texttt{Snapper}, when the \texttt{GetState} API is called by a transaction $T$, the \texttt{TransactionalActor} transparently controls when $T$ can get access to the actor state. More specifically, \texttt{Snapper} first identifies if $T$ is a PACT or an ACT by using the \texttt{TxnContext} information. Then, based on the different concurrency control strategies applied for PACT and ACT, \texttt{Snapper} inserts $T$ into the actor's local schedule and waits for the turn to execute $T$, i.e. grant $T$ access to the actor state. In addition, the \texttt{CallActor} API must be used when invoking a transactional call from one actor to another. This API acquires the \texttt{TxnContext} as one of the input parameters as well, so as to ensure the actor method invocations are executed under the same transaction context. 

\subsection{Integration of \texttt{SmSa} and \texttt{Snapper}}
In this section, we explain how we advance \texttt{Snapper}'s transactional layer to account for \texttt{SmSa} and achieve transactional guarantees on dependencies that cut across actors.
% To integrate \texttt{SmSa}, our rich actor state management layer, 
We start by removing from \texttt{Snapper} the opaque state management inherited by Orleans.
% two things need to be considered. First, the actor state should be presented as a \texttt{DictionaryState} object. Second, the dependency management tasks should be carried out as transactions.
As actor states are stored as a single object of the generic type, \texttt{TState}, we switch to a \texttt{SmSa}-managed map-based data structure, referred to \texttt{Dictionary-} \texttt{State} along this text. This prevents developers from creating arbitrary types to represent their application states for each actor, jeopardizing application design~\cite{quote_intro_2,quote_intro_3}.
% which often leads to an unnatural modeling of application entities.
% , jeopardizing the expressiveness of application entities .
In this mode, \texttt{SmSa} allows developers to manage entities through keys and associated actor references via the \texttt{DictionaryState} and its corresponding APIs (Get, Put, and Delete).
% In \texttt{Snapper}, the actor state is stored as a single object of the user-defined type -- \texttt{TState}, and the \texttt{GetState} API returns a reference to this object. To merge the data model of \texttt{SmSa} into \texttt{Snapper}, we just need to replace the \texttt{TState} object with the \texttt{DictionaryState} object, and any transactional operations that belong to $T$ should be performed on the \texttt{DictionaryState} via its external APIs (Get, Put, and Delete).

% user declare dependencies across actors based on managed record keys and Snapper must transparently, holisticallymanage declared dependices. For that, it must also  take into account declared dependices on scheduling transactions cutting across data-dependent actors.

The management of data dependencies in \texttt{SmSa} comprises two parts: (1) the registration and de-registration of dependencies and (2) the enforcement of dependencies. To manage dependency constraints transactionally, \texttt{SmSa} needs to rely on \texttt{Snapper}'s transactional APIs on certain occasions. First, when registering or a dependency, as is discussed in $\S$~\ref{sec:register-dependency}, the follower actor invokes the $RegisterDependency$ API, which in turn invokes a call to the leader actor via the \texttt{CallActor} API. Since the dependency list is stored in the \texttt{DictionaryState} object, adding or removing the dependencies to or from the list must occur by accessing the \texttt{DictionaryState} object via the \texttt{GetState} API. Second, \texttt{SmSa} enforces dependencies by resolving and forwarding operation logs to relevant actors; each will apply the forwarded logs and perform corresponding updates to the keys stored locally. Again, keys can only be accessed on the condition of getting access to the \texttt{DictionaryState} object by calling the \texttt{GetState} API.

\subsubsection{Workflow}
Fig.\ref{fig:workflow}b presents the workflow of a transaction in \texttt{SmSa-X} -- the  \texttt{Snapper} integrated with the advancements of \texttt{SmSa}. The figure marks the differences as red text compared to \texttt{Snapper} and uses the red arrows to represent steps that are relevant to dependency management. Step 8 creates a \texttt{DictionaryState} object for the transaction. \texttt{SmSa-X} does not allow transactions to access the original version of the actor state directly. Instead, a new \texttt{Dictionary-} \texttt{State} object is created for every transaction so as to isolate their read/write sets. In step 10, the transaction accesses the \texttt{DictionaryState} via its external APIs. When the actor finishes executing the user method, the operation logs are scanned to resolve the list of logs to forward (step 12) in order to enforce user-declared dependencies. Then, an \texttt{ApplyLogs} method is invoked on each dependent actor via the \texttt{CallActor} API. 

The actor who receives this method invocation will identify it as an internal method (step 13), which means it is implemented in the system code. Even if it is an internal method, it still needs to be executed under the transaction context; thus, a reference to the \texttt{DictionaryState} object should be acquired via the \texttt{GetState} API (step 15). While executing the internal method, the system code has access to the internal APIs of the \texttt{DictionaryState} object (step 16), such as \texttt{ApplyLogs}, which will read through the forwarded logs and apply updates to keys according to the registered dependencies. When the actor finishes executing the internal method, again, the operation logs need to be analyzed (step 18) because there might be updates performed on keys that trigger more operations to forward to other actors. Afterwards, a transaction log is persisted (step 19), and the logged changes made by the transaction are applied to the actor state (step 20). Note that, the same as in \texttt{Snapper}, the log writing happens once for every ACT or every batch of PACTs. As for step 20, given that each transaction has only operated on its own \texttt{DictionaryState}, \texttt{SmSa-X} needs to apply the changes again to the actual actor state when an ACT is committed or a batch of PACT has completed. By doing so, the results of one transaction are made visible to subsequent transactions. In \texttt{Snapper}, this step is carried out by overwriting the whole actor state, incurring a higher overhead than \texttt{SmSa-X}. % only performs the actual changes. 

Dependency registration and de-registration are implemented as internal methods. Changes made to the dependencies, including adding and removing dependencies, also need to be recorded first, persisted in the log record, and reflected in the actual actor state afterward. Note that adding or removing dependencies does not have a further effect, like an update on a leader key.

\subsubsection{Incremental Logging}
\label{sec:logging}
In \texttt{Snapper}, the actor state is always logged as a single object due to the lack of information about how the state is changed. We call this logging method as snapshot. It can be inefficient when only a small part of the actor state is modified. Differently, \texttt{SmSa-X} keeps a record of each change made to keys or dependencies (see the $OperationLog$ format in Fig.\ref{fig:model}); thus, \texttt{SmSa-X} only needs to write an incremental log for each transaction, which can largely reduce the logging overhead.

%In actor systems, each actor's state is accessed and persisted as a single object, which can be inefficient when only a small part of the actor state is changed. This issue is also discussed in a previous work, Snapper \cite{Snapper}, where the persistence of the $Order$ actor becomes more and more expensive due to the increasing number of data entries. There exist methods like event sourcing to keep track of the changes made to the actor state

\subsubsection{Key-level Concurrency Control for PACT}
\begin{figure}[h]
    \centering
    \includegraphics[width=\linewidth]{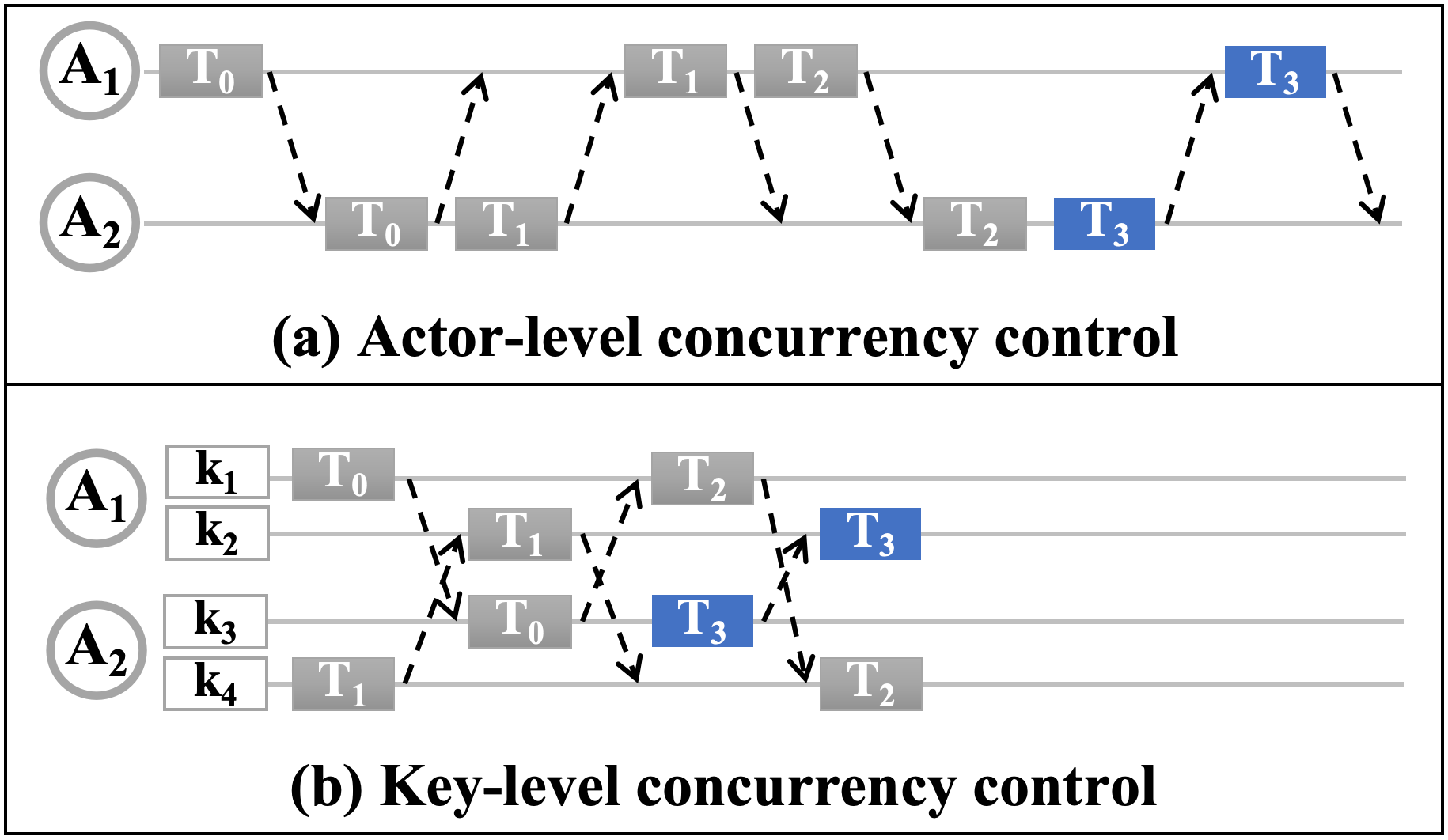}
    \caption{Granularity of concurrency control}
    \label{fig:key-cc}
\end{figure}

\texttt{Snapper} supports deterministic scheduling of transactions that provides the actor access information -- the set of actors to access and the number of times each actor will be accessed. As is listed in Fig.\ref{fig:workflow}a, the \texttt{StartTxn} API for PACT acquires an extra input parameter. \texttt{Snapper} uses this information to generate a deterministic transaction execution order for every related actor. As long as each actor executes PACTs in the ascending order of their transaction ID ($tid$), \texttt{Snapper} can guarantee global serializability. By integrating \texttt{SmSa}'s state management layer into \texttt{Snapper}, we devise a finer-grained transaction scheduling strategy. Given that \texttt{SmSa} acquires developers to specify which key to read or write in the application code, we can further assume that the set of keys accessed on each actor by the transaction is declared even before the transaction is started. Based on this key access information, a transaction execution schedule can be created for every single key. As is shown in Fig.\ref{fig:key-cc}, following a key-level schedule, a PACT on an actor only needs to wait for another PACT when the set of keys they access overlaps. 

Apparently, adopting key-level scheduling can achieve higher concurrency on each actor. However, it is not straightforward to implement it. First, a new interface should be created to receive transaction requests that tend to adopt key-level concurrency control. As is shown in Fig.\ref{fig:workflow}a, we introduce another \texttt{StartTxn} API that accepts the key access information as its input when a client submits the transaction request. Besides, the algorithm applied to generate the transaction execution schedule needs to be modified such that metadata is maintained for every single key. 

Furthermore, on each actor, it becomes insufficient to only check a PACT's \texttt{TxnContext} when the \texttt{GetState} API is called. A set of keys that the PACT tends to access must be given (Fig.\ref{fig:workflow}a). It is because the key-level schedule only specifies when it is safe for a PACT to access a specific key; therefore, the time when the \texttt{GetState} API returns the result varies depending on which key or a set of keys the PACT is acquiring. Besides, the PACT should only get access to the keys that it has acquired; otherwise, the transaction execution schedule may be violated. For example, suppose two PACTs $T_1$ and $T_2$ are scheduled on one actor. $T_1$ will only access the key $k_1$ and $T_2$ will access both $k_1$ and $k_2$. If $T_2$ acquires $k_2$ first, it can get the \texttt{DictionaryState} immediately. However, $T_2$ is not yet allowed to access $k_1$ because the status of $T_1$ is unknown. To safely access $k_1$, $T_2$ must call the \texttt{GetState} API with $k_1$ included in the key set. In \texttt{SmSa-X}, to accurately restrict the keys that a PACT can access, each \texttt{DictionaryState} object will only include the information of keys that the PACT has acquired. As illustrated in Fig.\ref{fig:workflow}b, in step 8, \texttt{SmSa-X} makes a copy of the key's information, including the key-value pair itself and its related dependencies. It also implies that to access the dependency information of a key, the transaction also needs to get access to the key first.

In addition, \texttt{SmSa-X} allows a transaction to call \texttt{GetState} API multiple times, each time may acquire the access right to different sets of keys. \texttt{SmSa-X} does not restrict the order to acquire different keys and the number of times each key is acquired. \texttt{SmSa-X} only maintains one \texttt{DictionaryState} instance for each transaction. When the \texttt{GetState} API is called for the first time, a new \texttt{DictionaryState} object is created for the transaction. Then, every time a new key is acquired via the \texttt{GetState} API, this key is added to the existing \texttt{DictionaryState} object. If the transaction asks for access to a key that already exists in the \texttt{DictionaryState}, no extra waiting is needed (e.g., steps 6, 7, 8 in Fig.\ref{fig:workflow}b). Note that the \texttt{GetState} API always returns the reference to the \texttt{DictionaryState} object that belongs to the current transaction, and developers can get access to this object or use this reference via any of the \texttt{GetState} calls. By doing so, a transaction can always access the latest actor state via the same reference, and the \texttt{GetState} API can be used for gradually extending the access right to different keys. 

An example is shown in Fig.\ref{fig:code} (lines 19-41). In this example, a \texttt{DeleteItemsInCart} transaction first gets access to its \texttt{DictionaryState} object by calling the \texttt{GetState} API without specifying any keys (line 21). Within this call, \texttt{SmSa-X} will create a new \texttt{DictionaryState} object for this transaction, which contains no keys. Afterward, the transaction can acquire access rights to a set of items (or keys) in two ways (lines 25-33, options 1 and 2). In both ways, the \texttt{state} object eventually gains the \texttt{RW} access to all acquired keys. In other words, keys are gradually added into the \texttt{DictionaryState} object and, therefore, can be accessed by the transaction. At last, the transaction deletes these items from the cart (lines 35-40).

As prescribed by \texttt{Snapper}, \texttt{SmSa-X} also requires for PACT the declaration of actors and keys accessed as part of a transaction, including the ones accessed during the dependency management process. As the example in Fig.\ref{fig:code}, each item in the cart actor is a replica of the corresponding product on the product actor (lines 3-17), which contains the latest product price. The replication here is expressed as an update dependency in \texttt{SmSa-X}. When items are deleted on the cart actor, it indicates the deletion of their dependencies stored on both the cart actor and the product actor (lines 35-40). Therefore, the transaction \texttt{DeleteItemsInCart} actually involves two actors and will access the same set of keys on each actor. The correct and complete key access information must be given to allow \texttt{SmSa-X} to schedule PACTs at the key level correctly.

\subsubsection{Key-level Concurrency Control for ACT}
\texttt{Snapper} also supports ACT, the type of transaction that does not declare the actor access information and applies non-deterministic execution. More specifically, each ACT gets access to an actor state by acquiring the RW lock maintained on the actor via the S2PL+wait-die protocol. To extend the actor-level concurrency control to the key-level concurrency control, \texttt{SmSa-X} can simply maintain a lock for every single key. Given that keys may be dynamically added and deleted on an actor, locks need to be added and removed accordingly. In \texttt{SmSa-X}, an ACT can add or delete a key $k$ on the condition that it gets the write lock of $k$ and a new lock for $k$ is created if it does not exist yet. Note that a key can only be deleted when there are no ACTs holding or waiting for the lock to avoid anomalies. 

\texttt{SmSa-X} provides interfaces for transactions to access keys individually while still preserving the ability to support actor-level concurrency control. The actor-level concurrency control is useful for cases where a transaction needs to scan the whole key-value collection on an actor or to query keys with certain predicates. \texttt{SmSa-X} allows developers to configure an actor to apply either actor-level or key-level concurrency control when it is created.
\section{Evaluation}
In this section, we conduct an extensive range of experiments to investigate the features of \texttt{SmSa} under various workloads. The first part of the experiments ($\S$~\ref{sec:exp-nt}, \ref{sec:exp-skew}, \ref{sec:exp-scale}) explores the characteristics of the basic building blocks of the data model. More specifically, we focus on the trade-offs of maintaining fine-grained key-value actor states and transaction processing performance. In the second part ($\S$~\ref{sec:exp-marketplace}), we turn our attention to popular cross-microservice correctness criteria sought by developers in practice~\cite{vldb2021}. In particular, we adopt the \texttt{Online Marketplace} benchmark and target exploiting the overhead of enforcing constraints cutting actors across, including foreign keys, data replication, and functional dependencies.

\subsection{Implementation Variants}
\begin{figure}[ht]
    \centering
    \includegraphics[width=\linewidth]{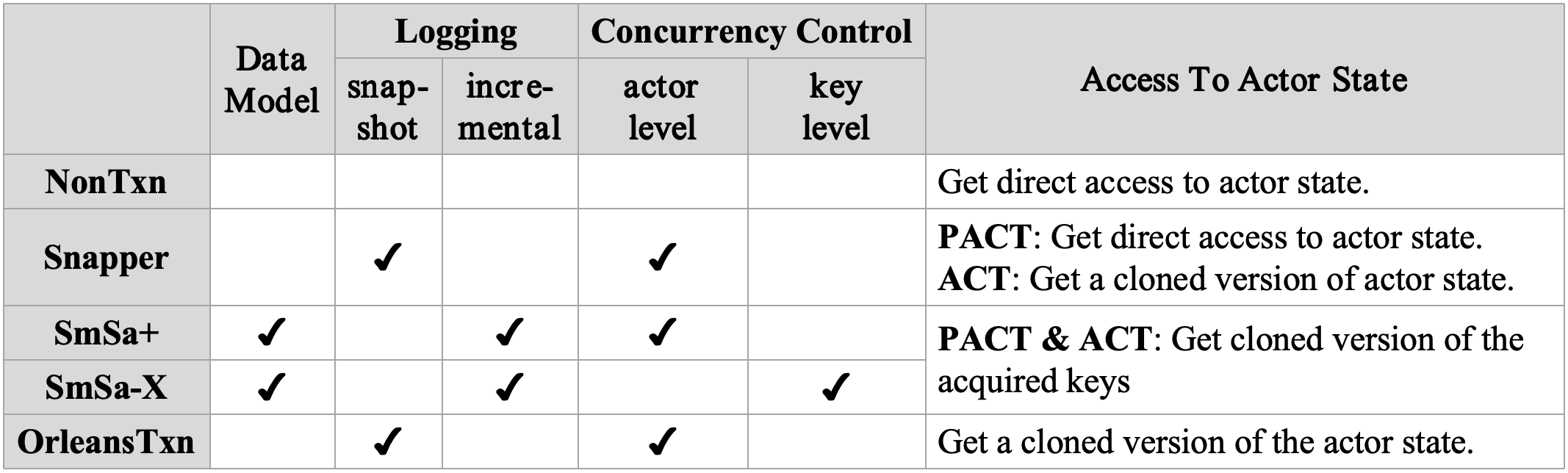}
    \caption{Implementation variants}
    \label{fig:variants}
\end{figure}
Five competing systems (Fig.\ref{fig:variants}) are compared to show the implications of our proposed advancements to performance. \textbf{\texttt{NonTxn}} applies a non-transactional execution on Orleans. Actors execute operations in arbitrary order; thus, accesses to the state are performed without isolation guarantees. It gives an upper bound of the system's performance, indicating the best throughput the actor system can achieve under a certain workload. \textbf{\texttt{Snapper}}~\cite{Snapper} is adopted in our experiments as a baseline solution for enforcing application correctness with strong consistency guarantees. To allow further insights, we integrate different data model functionalities in \texttt{Snapper} and devise two other variants. \textbf{\texttt{SmSa+}} combines the concept of keys and dependencies with \texttt{Snapper}, which facilitates transaction processing with incremental logging. \textbf{\texttt{SmSa-X}} further adds key-level concurrency control on top of \texttt{SmSa+}. Therefore, \texttt{SmSa-X} is the version that applies all the optimizations. The purpose of having \texttt{SmSa+} is to benchmark the effects of the different optimizations. In addition, we also include Orleans Transactions~\cite{OrleansTxn}(\textbf{\texttt{OrleansTxn}}) in our experiments. It is the default transaction management API in Orleans and applies 2PL and 2PC to fulfill ACID multi-actor transactions. That enriches the experiment by confronting competing concurrency control scheduling techniques, namely, locking-based in Orleans and deterministic in Snapper variants.

%\textbf{\texttt{SmSa+}} turns the actor states into a key-value table that is accessed with the data model's APIs, meanwhile preserving the transactional guarantees by adopting \texttt{Snapper}'s actor-level concurrency control method. Besides, with the data model, \texttt{SmSa+} is able to capture changes made on specific key-value pairs, thus facilitating fine-grained logging, which only persists the incremental changes, instead of persisting the whole actor state as is done in \texttt{Snapper}. Then, based on \texttt{SmSa+}, \textbf{\texttt{SmSa-X}} further exploits the information revealed by the data model and enables key-level concurrency control where transactions are scheduled according to the keys they access. \texttt{SmSa-X} has the potential to achieve higher concurrency than \texttt{Snapper} and \texttt{SmSa+}. 

\subsection{Experimental Setting}

\subsubsection{Deployment}
\label{sec:deploy}
All experiments are run on Orleans 8.1.0 and .NET SDK 8.0.300. Each experiment is conducted on a Orleans cluster consisting of a master node (\textbf{\texttt{MN}}) and several worker nodes (\textbf{\texttt{WN}}), each hosting a Orleans server and located in the same region. The \texttt{MN} is responsible for coordinating PACT execution on different \texttt{WN}s for \texttt{Snapper}, \texttt{SmSa+} and \texttt{SmSa-X}. In addition, a group of experiment nodes (\textbf{\texttt{EN}}) are spawned in the same region to generate workloads, host Orleans clients, and submit transaction requests to Orleans servers. The same number of \texttt{EN}s and \texttt{WN}s are deployed to ensure a sufficient amount of requests are generated and dispatched to \texttt{WN}s. Each node, regardless of the type, is an AWS EC2 instance (c5n) with a 2-core processor (4 vCPUs). In the scalability experiment ($\S$~\ref{sec:exp-scale} and \ref{sec:exp-marketplace}), we proportionally increase the number of \texttt{WN}s and \texttt{EN}s, as well as the number of vCPUs of the \texttt{MN}.

\begin{figure}[ht]
    \centering
    \includegraphics[width=0.8\linewidth]{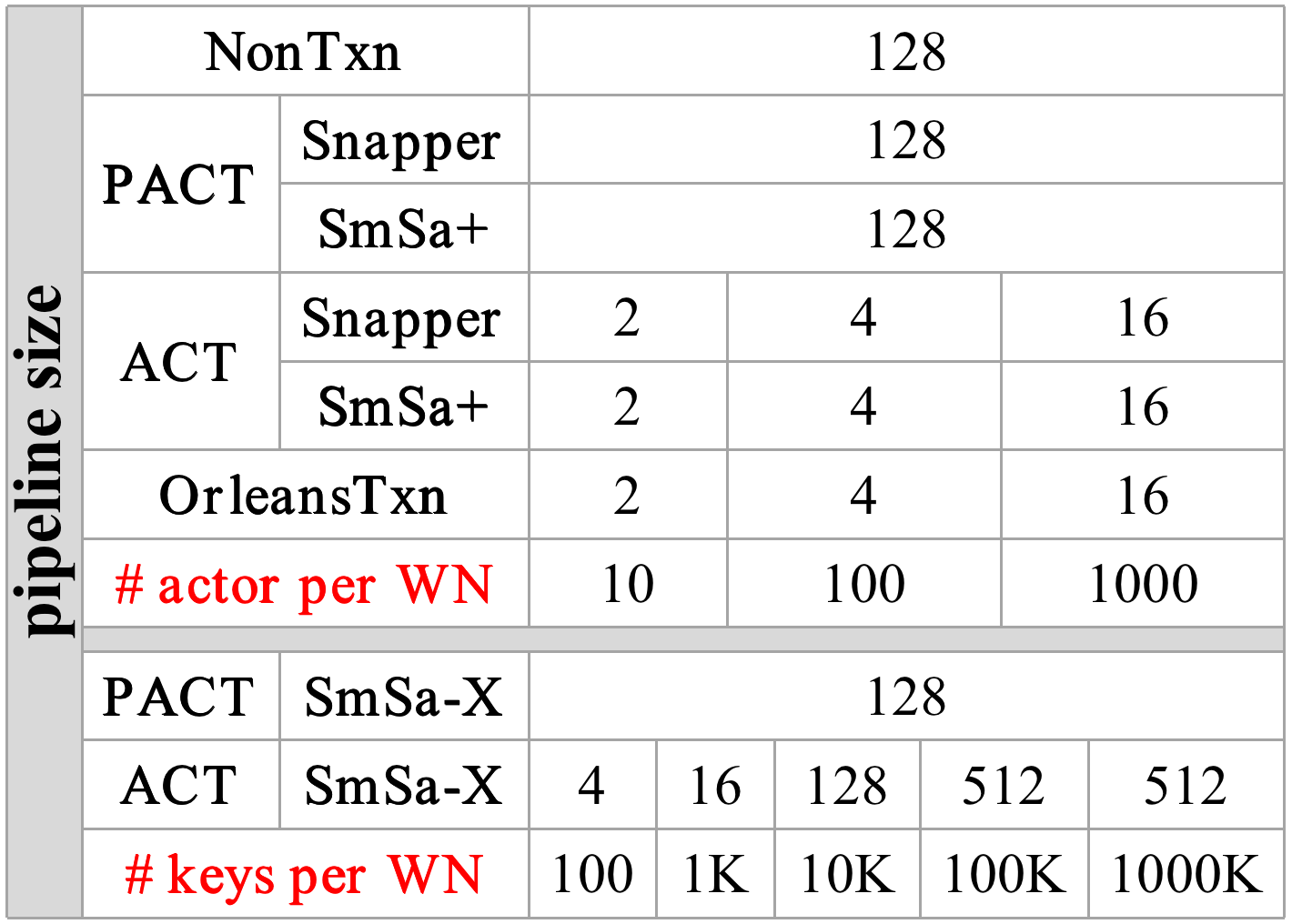}
    \caption{Pipeline size}
    \label{fig:pipe}
\end{figure}

On the client side, each \texttt{EN} initiates one Orleans client thread for PACT and ACT executions, respectively. The client thread submits a pipeline of transaction requests to \texttt{WN}s. The pipeline size determines the concurrency level of the workload. More specifically, it limits the maximum number of concurrent requests in the system. Whenever the result of one request is returned, a new request is replenished. In our experiments, the pipeline sizes are tuned for different implementation variants so that they all achieve a good performance while the system's computing resources are near saturation. Fig.\ref{fig:pipe} shows how the pipeline size is configured.

\subsubsection{\texttt{SmallBank} Benchmark}
\label{sec:smallbank}
We adopt the \texttt{SmallBank} ben- chmark~\cite{SmallBank} in the first part of the experiments to gain insights about the features of \texttt{SmSa}. In this benchmark, users' bank accounts are partitioned across many account actors, and operations such as \texttt{Deposit} and \texttt{Transfer} are applied to one or multiple account actors. \texttt{SmallBank} approximates an OLTP actor-oriented workload. Note that this benchmark shows no cross-actor dependencies, and we use it for investigating the performance of fundamental building blocks of \texttt{SmSa-X}, including incremental logging and the key-level concurrency control. In our experiments, we only employ the \texttt{MultiTransfer} transaction~\cite{ReactDB}, which withdraws money from accounts on one actor and deposits money to several accounts on other actors. This transaction gives us the flexibility to control the transaction size and the access pattern to different actors and accounts. More specifically, we adopt five parameters to configure a \texttt{SmallBank} workload. $numActor$ is the total number of account actors located in each \texttt{WN}. $actorSize$ determines the number of bank accounts, i.e., the number of keys, stored in each account actor. $txnSize$ specifies the number of keys a transaction will access on each selected actor. In the experiments, we fix the number of accessed actors for each transaction as 4. Thus, $txnSize = 4$ means the transaction will access a total of 16 keys across four actors. $actorSkew$ determines the number of hot actors on each \texttt{WN}. Each transaction selects a set of actors to access based on the following rule: there is a $75\%$ chance that an actor is selected from the set of hot actors. For example, when $numActor = 1000$ and $actorSkew = 1\%$, the hot set contains ten actors. And when $actorSkew = 100\%$, every actor has an equal chance to be chosen. The smaller $actorSkew$, the higher contention at the actor level. Similarly, $keySkew$ decides the number of hot keys on each actor and controls the contention at the key level.

\subsubsection{\texttt{Online Marketplace} Benchmark}
We also run experiments with the \texttt{Online Marketplace}~\cite{laigner2024benchmarking} benchmark, which mainly focuses on data management challenges in an event-driven and microservice-like system architecture. It simulates a multi-tenant web-scale application where sellers manage their products and associated inventory, and customers interact with the system by managing their carts (i.e., adding products) and submitting them for checkout. The reason for adopting this benchmark in our experiments is threefold. First, it covers several scenarios of cross-component data integrity constraints, which provides a perfect use case for our data model where the dependency between keys across actors is automatically handled. Existing commonly adopted benchmarks, such as YCSB~\cite{YCSB} and TPCC~\cite{TPCC}, do not model these types of correctness criteria, thus unfitting the goals of our experiments. Second, it can be easily mapped to the actor model by partitioning each component into multiple actors. Meanwhile, the application state can be easily modeled as key-value collections across different types of actors. Third, it forms a realistic and complex workload where some transactions can be implemented as PACT and some as ACT. Its business logic is carried out among eight different components as listed in Fig.\ref{fig:exp-marketplace}a. $\S$~\ref{sec:exp-marketplace} discussed more details about how the experiment is set up with this benchmark.

\begin{figure*}[h]
    \centering
    \includegraphics[width=\linewidth]{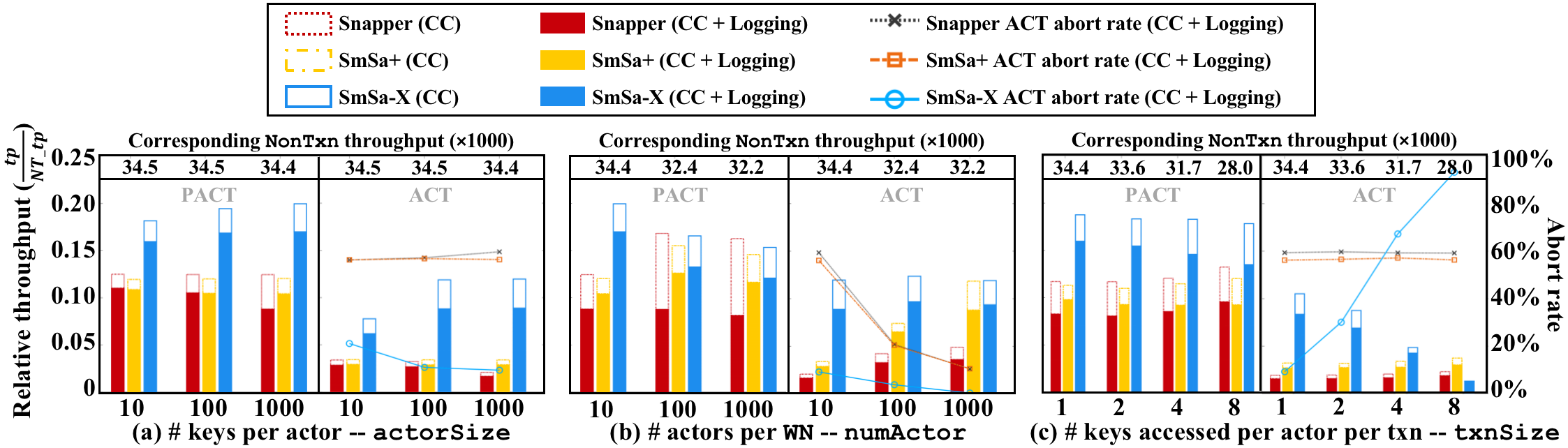}
    \caption{Characteristics of \texttt{SmSa}}
    \label{fig:exp-nt}
\end{figure*}

\subsection{Characteristics of \texttt{SmSa}}
\label{sec:exp-nt}
In this section, we investigate the impact of \texttt{SmSa} with the \texttt{SmallBank} benchmark. We measure the overhead of concurrency control and logging by presenting the relative throughput of \texttt{Snapper}, \texttt{SmSa+}, and \texttt{SmSa-X} with regards to the throughput of \texttt{NonTxn}. The differences between these three \texttt{Snapper} variants are supposed to reflect the trade-offs of a state management layer, namely, the overhead to maintain the key-value collection on each actor and the benefits it brings to the system performance. In this group of experiments, workloads are generated with a uniform distribution ($actorSkew = 100\%$ and $keySkew = 100\%$), and we vary the other three parameters, $numActor$, $actorSize$, and $txnSize$.

\subsubsection{Vary $actorSize$}
\label{sec:exp-actorSize}
In this experiment, we fix $numActor = 10$, $txnSize = 1$, and vary $actorSize$, Fig.\ref{fig:exp-nt}a shows the results. For PACT, when logging is disabled, the throughput of \texttt{Snapper} and \texttt{SmSa+} is not affected by $actorSize$, because they both apply actor-level concurrency control, in which the number of keys in an actor does not affect the contention at the actor level. Besides, the throughput of \texttt{SmSa+} is slightly lower than \texttt{Snapper}. It is because \texttt{Snapper} allows each PACT to access the actor state directly, while \texttt{SmSa+} needs to copy the accessed keys and operate on the cloned version. This is to fulfill the functionality of the data model that the before- and after-image of the modified keys are both captured. In contrast, when adding more keys to each actor, the throughput of \texttt{SmSa-X} increases because the key-level concurrency control benefits from the reduced contention on keys. And the gain from the finer-grained scheduling offsets the overhead of key cloning. When logging is enabled in PACT, \texttt{Snapper} throughput decreases largely, especially from $100$ to $1000$ keys per actor, while \texttt{SmSa+} and \texttt{SmSa-X} rarely change. This shows the advantage of incremental logging that \texttt{SmSa+} and \texttt{SmSa-X} only need to persist the changes on specific keys, but \texttt{Snapper} has to persist the whole actor state.

For ACT, \texttt{Snapper} and \texttt{SmSa+} show a similar trend as PACT -- their throughput and abort rate remains the same regardless of $actorSize$. Differently, \texttt{SmSa-X} throughput increases and the abort rate decreases. when $actorSize = 1000$, there is a significant gap between \texttt{Snapper} and \texttt{SmSa+}. This is because, in \texttt{Snapper}, every ACT has to make a copy of the whole actor state and apply read or write operations on the cloned state. Given that \texttt{Snapper} guarantees PACTs do not abort due to transaction conflicts, it is safe to have PACT modifying the actor state in place. However, every single ACT can be aborted; thus \texttt{Snapper} applies the updates of an ACT to the actor state only when the ACT commits.

\subsubsection{Vary $numActor$}
\label{sec:exp-numActor}
In this section, we vary $numActor$ and fix $txnSize = 1$, $actorSize = 1000$. Fig.\ref{fig:exp-nt}b shows the results. For PACT, the throughput of \texttt{Snapper} and \texttt{SmSa+} significantly increases from $10$ to $100$, then slightly decrease from $100$ to $1000$. When there are only $10$ actors in each \texttt{WN}, the contention on each actor is extremely high. In \texttt{Snapper} and \texttt{SmSa+}, each actor executes PACTs one by one in the ascending order of transaction ID ($tid$) and batch ID ($bid$). In addition, each actor is also responsible for tasks, including receiving batch messages and committing batches, which further extends the critical path of transaction processing. When more actors are added to the system, more transactions can be processed in parallel on different actors, therefore leading to a higher throughput. The decreased contention on actors also benefits their ACT execution -- the ACT throughout increases, and the abort rate decreases.

However, when we keep adding more actors, e.g. to $1000$ actors, the parallelism is not improved further due to the limited number of vCPUs. Meanwhile, the throughput drops when there are too many actors because the batching of PACTs becomes less efficient when the set of actors accessed by each PACT is less likely to overlap. We define the overlap rate $r$ as the number of PACTs included in a batch divided by the number of actors accessed by the batch. According to the collected experimental data, we get $r = 1.0, 0.4, 0.2$ for $numActor = 10, 100, 1000$ respectively. This overlap rate has a more obvious impact on \texttt{SmSa-X} that its throughput continuously decreases. In this experiment, even if more keys are added while adding more actors, the contention on keys remains low, so the performance of \texttt{SmSa-X} is not improved so much. The ACT abort rate of \texttt{SmSa-X} also keeps at a low value.

In terms of the difference between with and without logging, similar to the results observed in $\S$~\ref{sec:exp-actorSize}, the throughput of \texttt{Snapper} drops significantly when logging is enabled, such that its PACT throughput goes below \texttt{SmSa+} and \texttt{SmSa-X}.

\subsubsection{Effect of $txnSize$}
In this part, we fix $actorSize = 1000$, $numActor = 10$ and vary $txnSize$. Fig.\ref{fig:exp-nt}c shows the results. For \texttt{NonTxn}, its throughput obviously decreases with larger $txnSize$ because $txnSize$ determines the complexity of the transaction logic and the transaction execution latency. For both PACT and ACT, the absolute throughput of \texttt{Snapper} and \texttt{SmSa+} decreases, but the relative throughput slightly increases when $txnSize$ grows. It indicates they are not as sensitive to the change of $txnSize$ as \texttt{NonTxn}. PACT is mainly affected by actor-level transaction scheduling. Each PACT spends $\sim80\%$ time waiting for the turn to start execution on the first actor, and only $3\%$ time executing the transaction logic. For ACT, the dominant factors are the concurrency control (2PL) and the commit protocol (2PC). In addition, since both \texttt{Snapper} and \texttt{SmSa+} apply actor-level concurrency control, their abort rate remains the same even if $txnSize$ grows.

In contrast, \texttt{SmSa-X} is a lot more sensitive to the change of $txnSize$ compared to \texttt{Snapper} and \texttt{SmSa-X}. Its PACT throughput decreases due to the increased overhead of generating and maintaining the key-level transaction execution schedules, as well as persisting key modifications. The ACT throughput drops significantly, and the abort rate increases largely. In \texttt{SmSa-X}, an actor maintains a lock instance for every key stored on the actor, and each ACT needs to acquire the corresponding lock for every accessed key. Compared to PACT, ACT suffers more from contention. When $txnSize = 8$, the ACT throughput of \texttt{SmSa-X} drops below \texttt{Snapper} and \texttt{SmSa+}. For one thing, the number of concurrent transactions submitted to \texttt{SmSa-X} is much higher than \texttt{Snapper} and \texttt{SmSa+} ($128$ vs $2$). As is explained in $\S$~\ref{sec:deploy}, for \texttt{SmSa-X}, this number is tuned based on the total number of keys in each \texttt{WN}, while for \texttt{Snapper} and \texttt{SmSa+}, it is based on the total number of actors. In this experiment, when $txnSize$ grows from $1$ to $8$, the contention at the key level largely increases; thus, the ACT throughput of \texttt{SmSa-X} experienced a significant drop.

\subsubsection{Conclusion}
In conclusion, the data model is most effective when there are a small number of large actors, and the data model benefits more for workloads that access fewer keys. First, the data model helps the system capture changes performed on specific keys, thus largely reducing the logging overhead for \texttt{SmSa+} and \texttt{SmSa-X}. Second, the data model can further exploit concurrency in every single actor; therefore, \texttt{SmSa-X} performs significantly better than the other two when the contention is high at the actor level and low at the key level. Third, the overhead brought by the data model, such as the key cloning and the key-level transaction schedule maintenance, can be completely offset by the benefits brought by the finer-grained logging and higher concurrency level.

\begin{figure*}[h]
    \centering
    \includegraphics[width=\linewidth]{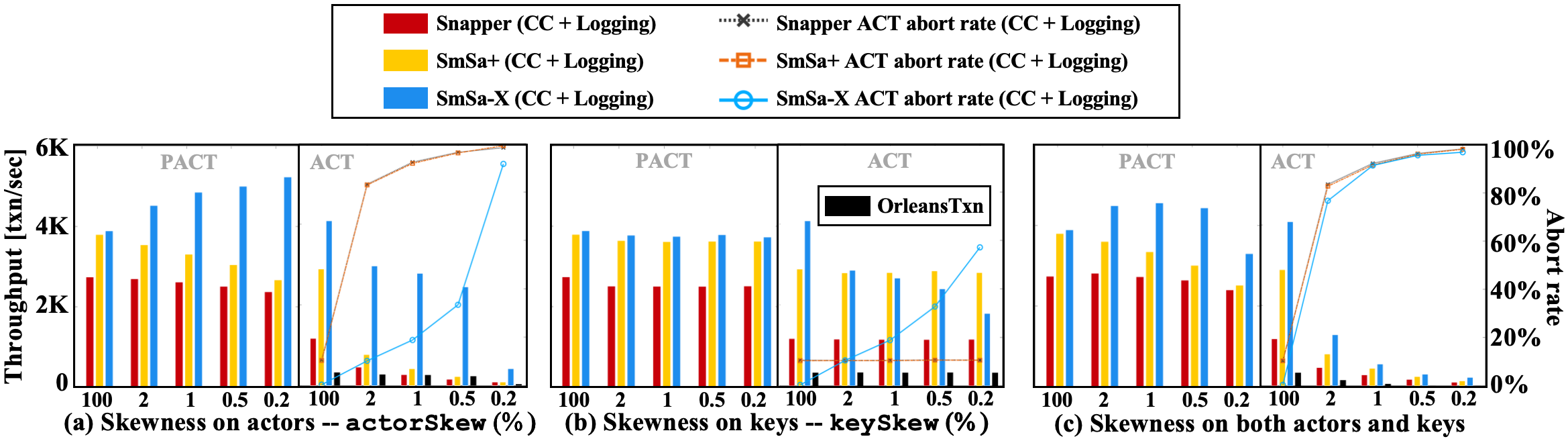}
    \caption{Effect of skewed workload (throughput)}
    \label{fig:exp-skew-tp}
\end{figure*}

\begin{figure*}[h]
    \centering
    \includegraphics[width=\linewidth]{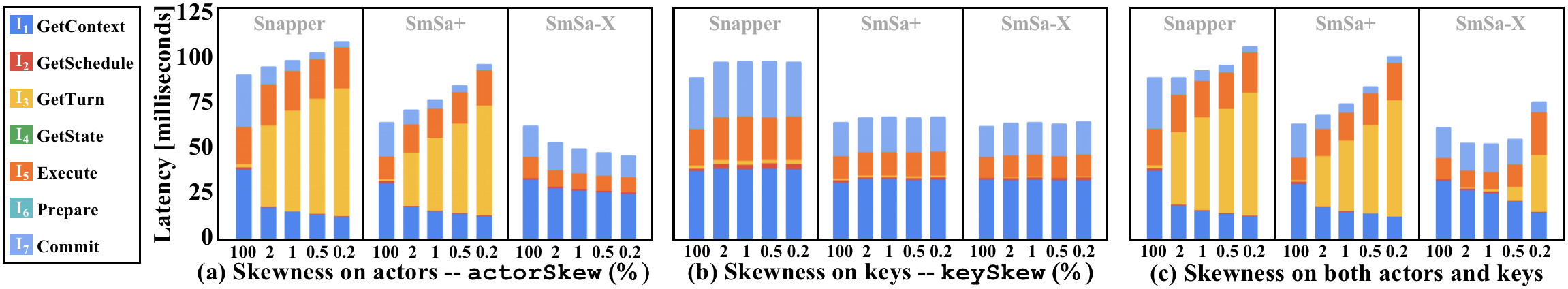}
    \caption{Effect of skewed workload (breakdown latency of PACT)}
    \label{fig:exp-skew-latency}
\end{figure*}

\subsection{Skewed workload}
\label{sec:exp-skew}
This section presents how performance is affected by skewed workloads. We fix $numActor = 1000$, $actorSize = 1000$, $txnSize = 1$ and vary $actorSkew$,  $keySkew$. Logging is enabled for all implementation variants from now onwards. In this section, we present not only the throughput of PACT and ACT (Fig.\ref{fig:exp-skew-tp}) but also the breakdown latency of PACT (Fig.\ref{fig:exp-skew-latency}). We divide the latency of each PACT into 7 time intervals. The breakpoints are set according to the progress of a PACT on the first accessed actor. The time intervals $I_1$, $I_2$, $I_3$, $I_4$, $I_6$ and $I_7$ represent the time spent on steps 2, 6, 7, 8, 21 and 23 in Fig.\ref{fig:workflow}b, respectively. Note that $I_5$ begins when $A$ starts to execute the transaction logic and ends when the whole transaction completes. $I_5$ includes the time to forward calls to other actors and execute these transactional invocations on other actors.

\subsubsection{Skew on actors}
\label{sec:exp-skew-actor}
Here, we investigate the impact of $actorSkew$, while fixing $keySkew = 100\%$. As explained in $\S$~\ref{sec:smallbank}, a smaller $actorSkew$ indicates higher contention at the actor level. Fig.\ref{fig:exp-skew-tp}a and \ref{fig:exp-skew-latency}a show the throughput and breakdown latency, respectively. For PACT, when $actorSkew$ decreases, \texttt{SmSa-X} benefits greatly from the growing contention on actors. As is discussed in $\S$~\ref{sec:exp-numActor}, when the workload is concentrated on a smaller set of actors, the batching of PACTs becomes more efficient due to a higher overlap rate. On the contrary, the contention on actors has a negative impact on \texttt{Snapper} and \texttt{SmSa+}. According to their breakdown latency, $I_3$ of \texttt{Snapper} and \texttt{SmSa+} grows obviously with decreasing $actorSkew$, indicating that each PACT is blocked for a longer time while waiting for previous PACTs to complete. Even if \texttt{Snapper} and \texttt{SmSa+} also benefit from the higher batching efficiency ($I_1$, $I_2$, $I_6$ and $I_7$ decrease), the increased $I_3$ dominates the transaction latency. Therefore, their throughput shows a decreasing trend. $I_3$ of \texttt{SmSa-X} remains very low because the contention on keys is low, and a PACT only needs to wait for the completion of a previous PACT that accesses the same key.

For ACT, the throughput of all three \texttt{Snapper} variants decreases. \texttt{Snapper} and \texttt{SmSa+} are affected by the contention at the actor level. However, \texttt{SmSa-X} is also affected by $actorSkew$ because it indirectly influences the contention at the key level. In addition, we measure the throughput of \texttt{OrleansTxn}. Given that \texttt{OrleansTxn} is reported to be extremely vulnerable to contention \cite{Snapper, OrleansTxnPerformance, OrleansTxnPerformance2, OrleansTxnGit}, here we use a workload with no deadlocks but under the same actor and key distribution ($actorSkew$ and $keySkew$). Deadlocks are removed by letting each transaction access the selected actors always in the ascending order of actor IDs. In our result, \texttt{OrleansTxn} remains a very low throughput under different $actorSkew$ values.

\subsubsection{Skew on keys}
\label{sec:exp-skew-key}
In this part, we present the impact of $keySkew$, which determines the contention at the key level. We fix $actorSkew$ as $100\%$. Fig.\ref{fig:exp-skew-tp}b and \ref{fig:exp-skew-latency}b show the throughput and breakdown latency, respectively. The PACT and ACT throughput of \texttt{Snapper} and \texttt{SmSa+}  rarely change while varying $keySkew$, because the contention on keys does not affect how transactions are scheduled at the actor level. For \texttt{SmSa-X}, its PACT throughput remains unchanged. The growing contention on keys does not cause a growing blocking time ($I_3$) for \texttt{SmSa-X} because the skewness is not high enough with $actorSkew = 100\%$. In contrast, its ACT throughput decreases largely. This validates that the ACT execution of \texttt{SmSa-X} is more sensitive to contention on keys than PACT. Again, \texttt{OrleansTxn} has a much lower throughput than the ACT execution of all three variants.

\begin{figure*}[h]
    \centering
    \includegraphics[width=\linewidth]{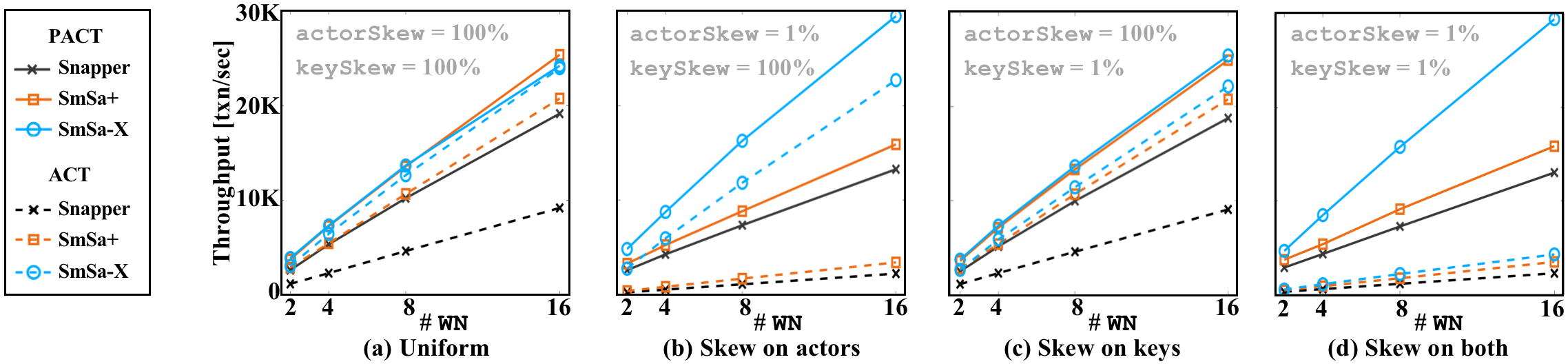}
    \caption{Scalability of \texttt{SmallBank}}
    \label{fig:exp-scale}
\end{figure*}
\subsubsection{Skew on both actors and keys}
Here, we present the combined effects of $actorSkew$ and $keySkew$ by changing their values from $100\%$ to $0.2\%$ simultaneously. Fig.\ref{fig:exp-skew-tp}c and \ref{fig:exp-skew-latency}c show the result. For \texttt{Snapper} and \texttt{SmSa+}, their PACT and ACT throughput show the same pattern as in $\S$~\ref{sec:exp-skew-actor} that they are not affected by contention on keys. For \texttt{SmSa-X}, its PACT throughput increases, then decreases. Compared to Fig.\ref{fig:exp-skew-tp}b, the contention on keys grows much faster in Fig.\ref{fig:exp-skew-tp}c. $I_3$ of \texttt{SmSa-X} also increases a lot from $1\%$ to $0.2\%$. For its ACT execution, the throughput has dropped significantly already from $100\%$ to $2\%$ because ACT is more sensitive to contention.

\subsubsection{Conclusion}
In conclusion, the PACT execution of \texttt{SmSa-X} greatly benefits from skewness on actors because of the improved batching efficiency. When there are a large number of actors in the system and the workload follows a uniform distribution, \texttt{SmSa-X} does not have an obvious advantage over \texttt{Snapper} and \texttt{SmSa+}. When the workload is skewed on actors, \texttt{SmSa-X} outperforms \texttt{SmSa+} and \texttt{Snapper}. In addition, the PACT execution of \texttt{SmSa-X} is less sensitive to contention on keys than its ACT execution. Its PACT throughput is affected only when the skewness on keys is at a very high level. the ACT execution of \texttt{SmSa-X} performs better than \texttt{Snapper} and \texttt{SmSa+} in most cases, except when the contention on keys is extremely high.

\begin{figure*}[h]
    \centering
    \includegraphics[width=\linewidth
    ]{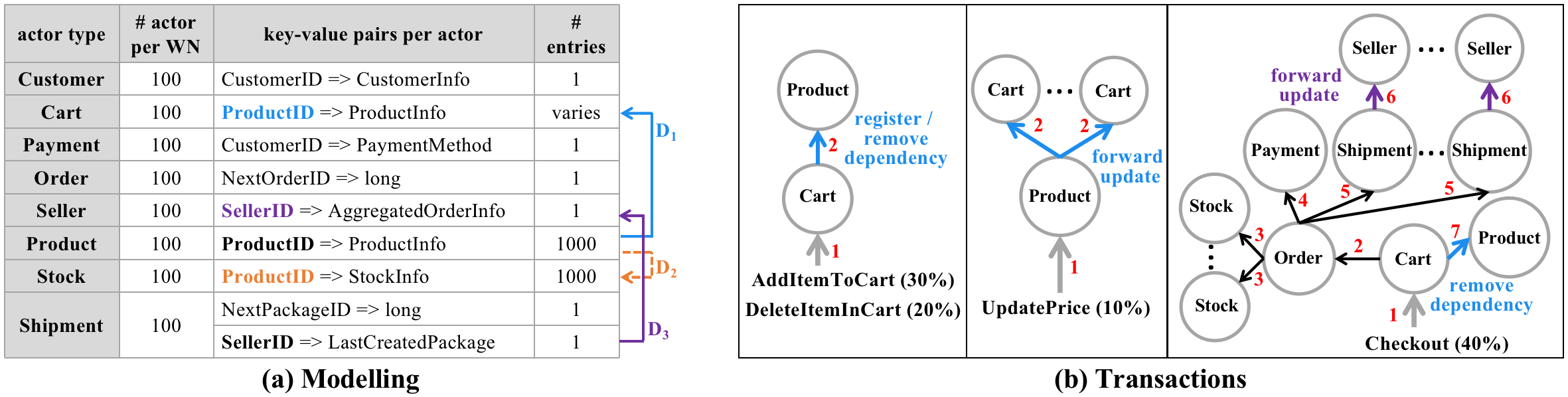}
    \caption{\texttt{Online Marketplace} experimental setting}
    \label{fig:marketplace}
\end{figure*}

\subsection{Scalability}
\label{sec:exp-scale}
In this section, we validate the scalability of \texttt{SmSa}. We measure the throughput of PACT and ACT of three variants (\texttt{Snapper}, \texttt{SmSa+}, and \texttt{SmSa-X}) under different $actorSkew$ and $keySkew$, with $numActor = 1000$, $actorSize = 1000$, $txnSize = 1$. Fig.\ref{fig:exp-scale} shows that all cases scale linearly. The same as in $\S$~\ref{sec:exp-skew}, \texttt{Snapper} and \texttt{SmSa+} are only affected by $actorSkew$, not $keySkew$. When $actorSkew$ changes from $100\%$ to $1\%$, the contention at the actor level increases. The PACT throughput of \texttt{SmSa+} decreases from $25K$ to $16K$, and \texttt{Snapper} from $19K$ to $13K$. The ACT throughput of \texttt{SmSa+} decreases from $20K$ to $3K$, and \texttt{Snapper} from $9K$ to $2K$. Their ACT throughput decreases more significantly than the PACT because ACT execution is more sensitive to contention at the actor level. Differently, for \texttt{SmSa-X}, its PACT throughput even increases from $25K$ to $30K$ when the contention on actors grows. And its ACT throughput only drops when the contention on keys is extremely high (Fig.\ref{fig:exp-scale}d).

\begin{figure*}[h]
    \centering
    \includegraphics[width=\linewidth
    ]{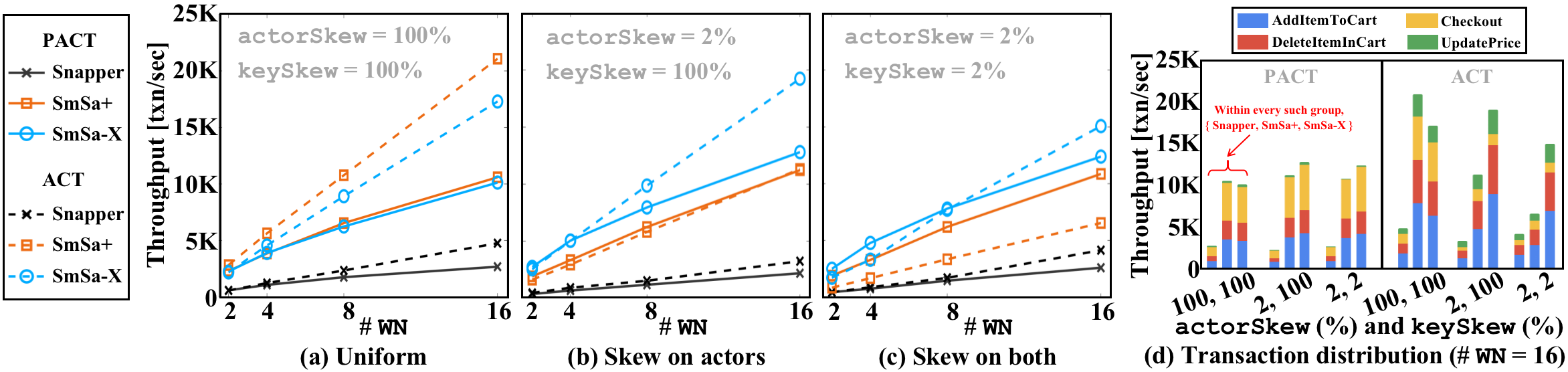}
    \caption{Performance of \texttt{Oneline MarketPlace}}
    \label{fig:exp-marketplace}
\end{figure*}

\subsection{Online Marketplace}
\label{sec:exp-marketplace}
In this section, we measure the performance of different implementation variants under \texttt{Online Marketplace} benchmark~\cite{laigner2024benchmarking}.
%, a benchmark that models important complex cross-component relationships that arise in real-world distributed applications. In particular, \texttt{Online Marketplace} prescribes data integrity constraints that include foreign keys, data replication consistency, and functional dependencies. 
For \texttt{Snapper}, we adopt conventional actor state manipulation and encode all dependencies in the application logic. \texttt{SmSa+} benefits from the fine-grained state manipulation, which ought to decrease logging overhead. For \texttt{SmSa-X}, we exploit the full contributions of this work, providing a sophisticated actor state manipulation, transparent constraint enforcement, and novel logging and concurrency control techniques.

\subsubsection{Modelling}
The \texttt{Online Marketplace} contains eight different components, each encapsulating corresponding application logic and maintaining a set of relations. In our implementation, as is shown in Fig.\ref{fig:marketplace}a, each component is mapped to a group of actors. Each seller maintains $1000$ products, and $100$ sellers make a total of $100K$ products in the system. Besides, three dependency constraints are modelled. First, each item in a cart actor is essentially a replica of the product in the corresponding product actor; thus, an update dependency is built between the original and the copied keys. Second, the stock of a product in the stock actor is a foreign key of the product in the product actor, which can be interpreted as a delete dependency. Third, each seller actor maintains an up-to-date materialized view of the total number of orders created. We implement it by using a functional update dependency.

\subsubsection{Transactions}
In our experiments, four types of transactions are adopted (Fig.\ref{fig:marketplace}b). The \texttt{Checkout} transaction crosses seven types of actors, and the total number of actors involved depends on the number of items that are bought. Steps 6 and 7 represent tasks that happen while resolving the dependency constraints. In our experiment, a workload consists of $30\%$ \texttt{AddItemToCart}, $20\%$ \texttt{DeleteItemInCart}, $10\%$ \texttt{Update-} \texttt{Price} and $40\%$ \texttt{Checkout} transactions. Each transaction is generated by selecting a customer or a seller under the $actorSkew$ and selecting a product under the $keySkew$. For a \texttt{Checkout} transaction, the number of items to checkout varies from 1 to 5. In addition, based on the observation from previous experiments that ACT is vulnerable to contention, we implement transactions as PACT whenever we can in this experiment. Therefore, only \texttt{UpdatePrice} transaction is executed as ACT because it is unknown which cart actors have dependencies on the product when a \texttt{UpdatePrice} transaction is submitted.

\subsubsection{Scalability under Skewed Workloads}
Fig.\ref{fig:exp-marketplace}a shows the result of a scalability experiment on \texttt{Online Marketplace} workload. Fig.\ref{fig:exp-marketplace}b shows the throughput of each type of transaction when $\#WN = 16$. Here, the PACT throughput represents the deterministic execution of all transactions, except for \texttt{UpdatePrice}.

For PACT, under the uniform distribution, \texttt{SmSa+} throughput is slightly higher than \texttt{SmSa-X}. When $actorSkew$ decreases to $2\%$, their PACT throughput both increases; however, \texttt{SmSa-X} surpasses \texttt{SmSa+}. It is because they both benefit from higher batching efficiency, but \texttt{SmSa+} is further affected by the increased blocking on each actor, while \texttt{SmSa-X} remains at a higher concurrency level, benefiting from key-level concurrency control. When $keySkew$ is decreased to $2\%$, the throughput of \texttt{Snapepr+} and \texttt{SmSa-X} both decrease moderately, and the impact on \texttt{UpdatePrice} transaction is more severe than on other transactions.

The ACT throughput of all three variants is higher than the corresponding PACT throughput. As is shown in Fig.\ref{fig:exp-marketplace}b, the transaction distribution of ACT differs from PACT -- the throughput of smaller transactions such as \texttt{AddItemToCart} and \texttt{DeleteItemInCart} accounts for a much higher proportion in ACT execution. This is because ACT has the flexibility to quickly abort large transactions, in exchange for the commit of smaller-sized transactions. For \texttt{SmSa-X}, its ACT throughput increases when $actorSkew$ changes to $2\%$, because it gains throughput from smaller transactions by aborting larger transactions. When further changing $keySkew$ to $2\%$, its ACT throughput decreases. It is because the contention on keys becomes very high, such that smaller transactions also suffer from the high contention. Thus, all types of transactions experience a higher abort rate and result in a lower throughput. 

\begin{figure}[ht]
    \centering
    \includegraphics[width=\linewidth]{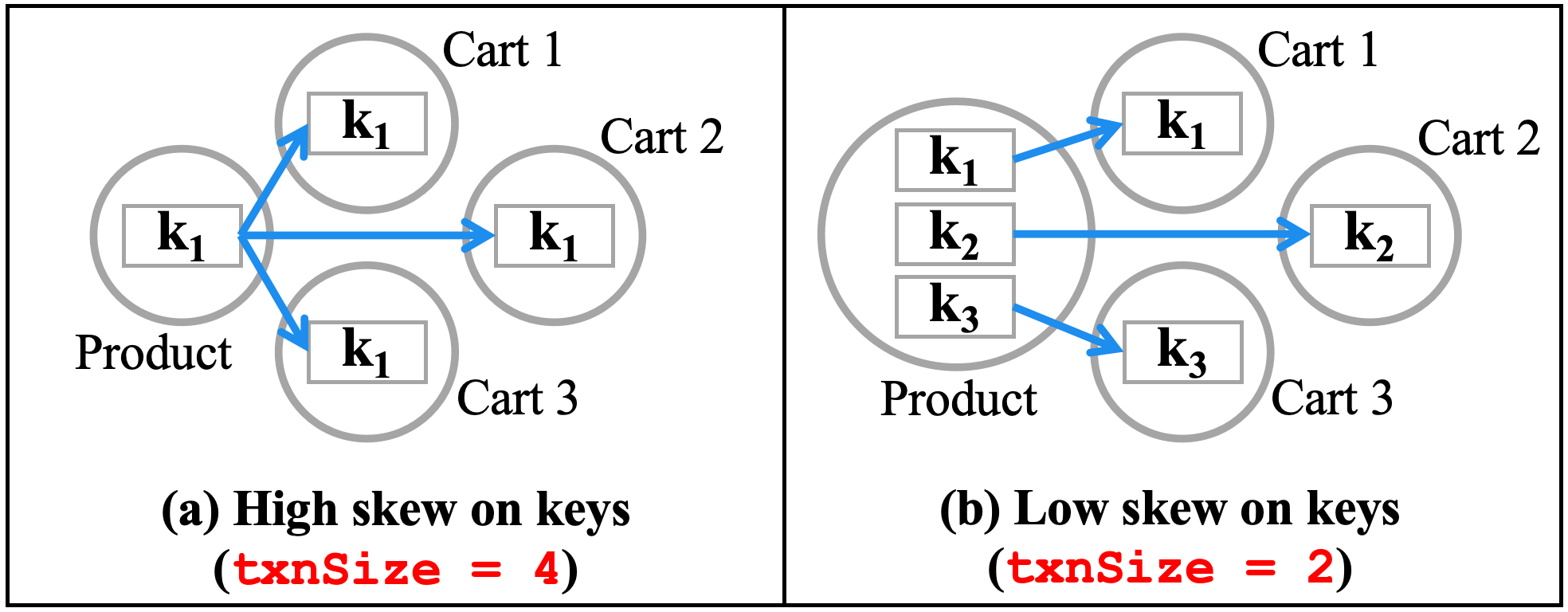}
    \caption{Different skewness on keys}
    \label{fig:keyskew}
\end{figure}

For \texttt{SmSa+}, its ACT throughput decreases from $21K$ to $11K$ when $actorSkew$ changes from $100\%$ to $2\%$. It is because \texttt{SmSa+} suffers from contention on actors. However, when further varying $keySkew$ from $100\%$ to $2\%$, \texttt{SmSa+} throughput drops from $11K$ to $6K$. Note that when $keySkew$ decreases, it raises the possibility that the same product is added by many carts and further causes a larger \texttt{UpdatePrice} transaction (Fig.\ref{fig:keyskew}), therefore bringing higher contention at the actor level. As for \texttt{Snapper}, its PACT and ACT throughput remain at a low level in all three groups due to the high logging overhead.

In conclusion, under the \texttt{Online Marketplace} benchmark, where the dependency constraints are broadly applied, \texttt{SmSa-X} shows its advantage over the other two variants in both PACT and ACT execution, and it reacts to different skewed workloads in a similar way as in $\S$~\ref{sec:exp-skew}.
\section{Related Work}

Data replication is a mechanism widely adopted in distributed systems to leverage data locality, decrease data access latency, enhance fault tolerance, and achieve higher system availability. Existing actor systems adopt methods like event sourcing \cite{Akka, Orleans} and geo-distributed caching \cite{GEO} to replicate the actor state, achieving eventual consistency and linearizability, respectively. However, actor replication differs from replicating data items, since the technique involves actor metadata (e.g., type), a possibly large actor state and actor functionalities, thus introducing a higher overhead. %In an actor-based application, a copy of a data item is used to co-locate with data on other actors and serve different purposes; thus, copying a whole actor does not tickle the real problem. 
In this work, an actor state abstraction is designed to facilitate the identification and replication of data items across actors while capturing and maintaining the primary-replica relations.

The actor system, Akka \cite{Akka}, relies on the conflict-free replicated data types (CRDTs) \cite{CRDT, Akka-CRDT} to replicate data across nodes with eventual consistency guarantee. Although not incurring the high overhead of actor replication, having data types as the abstraction level inherits the same impedance found in Orleans, updates to single entities are treated as an update to the whole actor state, preventing optimizations. As a result, such method is oblivious to the possible data dependencies across actors. %Nevertheless, it validates the feasibility of a data model where each data item is attached with a unique key and the practicality of the key-value-like API. 

Anna~\cite{anna} is a distributed key-value store that employs the actor model transparently to developers in order to process and possibly merge concurrent updates. Different from our work, Anna does not expose an actor programming model nor offer data dependency management and constraint enforcement. Besides, Anna focus on eventual consistency scenarios, contrasting with \texttt{SmSa} support for ACID guarantees. %and transactional constraint maintenance. % can also say: besides, Anna does not offer ACID
DPA~\cite{kraft} is a data platform for OLAP workloads where developers utilize an actor-based programming model to abstract and advance an underlying data analytics system. DPA is designed for bulk data updates, which inhibits its use in event-driven, highly-transactional scenarios as found in microservices.

% The replication of data also often takes the form of caching. In the context of actor systems, due to the share-nothing feature of the actor model, where each actor encapsulates its private state and stays isolated from other actors' memory space, a collection of actors forms a virtually distributed environment. Therefore, replicating data to different actors for better performance becomes a natural need and faces the same challenge as maintaining data consistency. 

Data dependency is another crucial aspect of data management, which specifies how data in one object depends on data in others. To enforce data dependency constraints in the context of actor systems, recent research works \cite{ActorDB, ReactDB, Wang2019} have focused on connecting actors to relational database systems so as to regain their support for declarative querying and extensive data management functionalities. In this approach, operations on data are forwarded through actors to the backend storage and handled there. However, this model conflicts with the function shipping paradigm, which decouples the fast computation from the slow storage \cite{AODB}. In contrast, dependencies between different data items across actors are captured by \texttt{SmSa} in the application layer. Meanwhile, dependency constraints are enforced with transactional guarantees.

%we find the necessity of a data model to identify data items, keep track of relations between data items, and preserve the constraints defined by the data dependencies.

%Nevertheless, there is an inevitable need to orchestrate and coordinate among multiple actors to correctly serve a request in various scenarios. 

%It refers to the relationships among different data, specifying how data in one object depends on data in others. The relationship between data has been thoroughly investigated in relational database systems, and data integrity is one of the most important properties of an application. However, as for the actor model, its design principle stands almost on the opposite side of the data dependency. An ideal and highly performant actor-based application expects each actor to proceed completely independently.
\section{Conclusion}

The actor model emerged as a promising concurrency model for facilitating distributed application design. However, exposing an opaque state management abstraction limits developers' ability to express complex relationships that cut across actors, inhibiting further adoption of the actor model.

To fill this gap, we propose \texttt{SmSa-X}, a state management layer that advances the actor model with rich state management abstractions and novel logging and concurrency control algorithms. Our experimental study shows \texttt{SmSa-X} enhances the design of complex relationships in actor systems and improves by 2X the performance of state-of-the-art deterministic concurrent control methods. As a result, \texttt{SmSa-X} will facilitate designing highly partitioned and distributed data-intensive applications based on the actor model. As for future work, we identify the potential to further increase \texttt{SmSa-X}'s performance when there is high contention on keys. Though there is little space to further exploit concurrency on an individual actor, techniques like actor re-balancing or key re-partitioning can be adopted.

\begin{comment}
future work: 

the data dependency management brings performance gains

target case: an update on one actor needs to be forwarded to multiple other actors, SnapperX can group a batch of such transaction together and only forward one message to the follower actor, instead of one message per transaction

more example: if multiple UpdateProduct happen in a row, can apply this optimization...

need to explain: dealock issue when batching updates together
eg. T1: A0 =====> A1
    T2: A0 <===== A1
if T1, T2 are within one batch, then the forward from A0 to A1 must happen when T1 is finished, instead of the whole batch finished ===> it is developers responsibility to manage such deadlocks....
\end{comment}

\begin{acks}
This work was supported by Independent Research Fund Denmark under Grant 9041-00368B.
\end{acks}

\bibliographystyle{ACM-Reference-Format}
\bibliography{paper}

%\appendix

\end{document}